\documentclass[secnumarabic, twocolumn, amssymb, floatfix, nobibnotes, superscriptaddress, aps, prmaterials]{revtex4-2}

\usepackage{etoolbox}
\usepackage{amsmath}
\usepackage{amsfonts}
\usepackage{graphicx}
\usepackage{braket}
\usepackage{booktabs}
\usepackage{bbm}
\usepackage{float}
\usepackage{nicefrac}
\usepackage{bm}
\usepackage{hhline}
\usepackage{multirow}
\usepackage{lipsum}
\usepackage[colorinlistoftodos,prependcaption,textsize=tiny]{todonotes}

\usepackage{hyperref}

\begin{document}

\title{An Optimally-Tuned Starting Point for Single-Shot \texorpdfstring{$GW$}{GW} Calculations of Solids}

\author{Stephen E. Gant}%
\affiliation{Department of Physics, University of California Berkeley, Berkeley, California 94720, USA}

\author{Jonah B. Haber}%
\affiliation{Department of Physics, University of California Berkeley, Berkeley, California 94720, USA}

\author{Marina R. Filip}%
\affiliation{Department of Physics, University of Oxford, Clarendon Laboratory, Oxford OX1 3PU, United Kingdom}

\author{Francisca Sagredo}%
\affiliation{Department of Physics, University of California Berkeley, Berkeley, California 94720, USA}

\author{Dahvyd Wing}%
\affiliation{Department of Molecular Chemistry and Materials Science, Weizmann Institute of Science, Rehovoth 76100, Israel}

\author{Guy Ohad}%
\affiliation{Department of Molecular Chemistry and Materials Science, Weizmann Institute of Science, Rehovoth 76100, Israel}

\author{Leeor Kronik}%
\affiliation{Department of Molecular Chemistry and Materials Science, Weizmann Institute of Science, Rehovoth 76100, Israel}

\author{Jeffrey B. Neaton}%
\email[Corresponding author: ]{jbneaton@lbl.gov}
\affiliation{Department of Physics, University of California Berkeley, Berkeley, California 94720, USA}
\affiliation{Materials Sciences Division, Lawrence Berkeley National Laboratory, Berkeley, California 94720, USA}
\affiliation{Kavli Energy NanoScience Institute at Berkeley, Berkeley, California 94720, USA}

\begin{abstract}
The dependence of \emph{ab initio} many-body perturbation theory within the $GW$ approximation on the eigensystem used in calculating quasiparticle corrections limits this method's predictive power. Here, we investigate the accuracy of the recently developed Wannier-localized optimally tuned screened range-separated hybrid (WOT-SRSH) functional as a generalized Kohn-Sham starting point for single-shot $GW$ ($G_0W_0$) calculations for a range of semiconductors and insulators. Comparison to calculations based on well-established functionals, namely, PBE, PBE0, and HSE, as well as to self-consistent $GW$ schemes and to experiment, shows that band gaps computed via $G_0W_0$@WOT-SRSH have a level of precision and accuracy that is comparable to that of more advanced methods such as quasiparticle self-consistent $GW$ and eigenvalue self-consistent $GW$. We also find that $G_0W_0$@WOT-SRSH improves the description of states deeper in the valence band manifold.
Finally, we show that $G_0W_0$@WOT-SRSH significantly reduces the sensitivity of computed band gaps to ambiguities in the underlying WOT-SRSH tuning procedure.
\end{abstract}

\maketitle

\section{Introduction}
\textit{Ab initio} many-body perturbation theory within the $GW$ approximation is a state-of-the-art approach for calculations of the quasiparticle (QP) band structures of crystalline solids  \cite{hedinNewMethodCalculating1965a,hedinEffectsElectronElectronElectronPhonon1970,
strinatiDynamicalCorrelationEffects1980,strinatiDynamicalAspectsCorrelation1982,
hybertsenFirstPrinciplesTheoryQuasiparticles1985,hybertsenElectronCorrelationSemiconductors1986,
godbyAccurateExchangeCorrelationPotential1986,aryasetiawanTheGWmethod1998d,
aulburQuasiparticleCalculationsSolids2000a,martinElectronicStructureBasic2020,
onidaElectronicExcitationsDensityfunctional2002d,brunevalBenchmarkingStartingPoints2013a,
faberExcitedStatesProperties2014a,martinInteractingElectronsTheory2016b,reiningGWApproximationContent2018,
golzeGWCompendiumPractical2019}.
In the $GW$ approximation, the self-energy $\Sigma$ is given by the convolution $\Sigma=iGW$, where $G$ is the single particle Green's function and $W$ is the dynamically screened Coulomb interaction. The $GW$ self-energy is normally first constructed from a (generalized) Kohn-Sham (GKS) \cite{seidlGeneralizedKohnShamSchemes1996a} ``starting point", an eigensystem computed from density functional theory (DFT). While semi-local functionals, such as the local density approximation \cite{hohenbergInhomogeneousElectronGas1964} or generalized gradient approximations like PBE \cite{perdewGeneralizedGradientApproximation1996},
have historically been the standard choice for constructing this starting point eigensystem \cite{hybertsenFirstPrinciplesTheoryQuasiparticles1985,aryasetiawanTheGWmethod1998d,aulburQuasiparticleCalculationsSolids2000a,onidaElectronicExcitationsDensityfunctional2002d}, hybrid functionals are increasingly used \cite{fuchsQuasiparticleBandStructure2007b,korzdorferStrategyFindingReliable2012,atallaHybridDensityFunctional2013,refaely-abramsonQuasiparticleSpectraNonempirical2012a,dauthPiecewiseLinearityGW2016a,knightAccurateIonizationPotentials2016,maromAccurateDescriptionElectronic2017a,leppertPredictiveBandGaps2019a,golzeGWCompendiumPractical2019,hellgrenElectronicStructureMathrmTiSe2021}.

In practice, there are a variety of choices regarding how $GW$ calculations are carried out, with significant consequences for accuracy \cite{rangelReproducibilityCalculationsSolids2020,golzeGWCompendiumPractical2019}. Once the $GW$ self-energy has been constructed, the quasiparticle energies can be computed via first-order correction to the GKS eigenvalues, the so-called single-shot $GW$ ($G_0W_0$) approach \cite{hybertsenFirstPrinciplesTheoryQuasiparticles1985}, or $G$ and/or $W$ can be iterated to self-consistency \cite{surhQuasiparticleEnergiesCubic1991,schoneSelfConsistentCalculationsQuasiparticle1998b,luoQuasiparticleBandStructure2002a,faleevAllElectronSelfConsistentGW2004,vanschilfgaardeQuasiparticleSelfConsistentGW2006b,shishkinSelfconsistentGWCalculations2007,kotaniQuasiparticleSelfconsistentGW2007,shishkinAccurateQuasiparticleSpectra2007,kutepovGroundstatePropertiesSimple2009a,korzdorferStrategyFindingReliable2012,refaely-abramsonGapRenormalizationMolecular2013a,eggerOutervalenceElectronSpectra2014b,atallaHybridDensityFunctional2013,skoneSelfconsistentHybridFunctional2014,gallandiLongRangeCorrectedDFT2015a,gallandiAccurateIonizationPotentials2016a,knightAccurateIonizationPotentials2016,jiangGWLinearizedAugmented2016,kutepovElectronicStructureNa2016,dauthPiecewiseLinearityGW2016a,carusoBenchmarkGWApproaches2016a,maromAccurateDescriptionElectronic2017a,boisSizeDependenceNonempiricallyTuned2017,kutepovSelfconsistentSolutionHedin2017b,grumetQuasiparticleApproximationFully2018,hellgrenElectronicStructureMathrmTiSe2021,liTuningRangeSeparation2021}. The $G_0W_0$ method is the computationally least expensive approach, and, as has been well established, the quasiparticle band structures computed with $G_0W_0$ approaches typically substantially improve agreement with experiment compared to those obtained directly from the GKS eigenvalues of their underlying DFT starting points \cite{fuchsQuasiparticleBandStructure2007b,chenAccurateBandGaps2015a, jiangGWLinearizedAugmented2016,grumetQuasiparticleApproximationFully2018,golzeGWCompendiumPractical2019}. For example, QP band gap data from an analysis of $G_0W_0$ calculations for various semiconductors and insulators by Grumet et al.\ \cite{grumetQuasiparticleApproximationFully2018} exhibited a mean absolute error (MAE) of $0.2$ eV compared to an MAE of $1.2$ eV for the underlying DFT functionals used. However, $G_0W_0$ results exhibit a starting-point dependence, where results can depend considerably on the DFT functional used to construct the starting eigensystem \cite{fuchsQuasiparticleBandStructure2007b,brunevalBenchmarkingStartingPoints2013a,vansettenAutomationMethodologiesLargescale2017,golzeGWCompendiumPractical2019,leppertPredictiveBandGaps2019a}.
For molecules, a range of about $1$ eV in $G_0W_0$ calculations of highest occupied molecular orbital energies has been reported \cite{maromBenchmarkGWMethods2012a,sharifzadehManybodyPerturbationTheory2018}. Likewise, $G_0W_0$ results for solids (e.g., Si, InN, ZnO, ZnS, CdS, and GaN) \cite{fuchsQuasiparticleBandStructure2007b,rinkeCombiningGWcalculationsExactexchangeDensityfunctional2005b} have shown a similar starting point dependence of up to $2$ eV in computed band gaps. As such, it is common practice to differentiate $G_0W_0$ calculations by the functional used in their starting point, denoted by $G_0W_0$@(...). Relatedly, the accuracy of $G_0W_0$ calculations based on semi-local DFT functionals is known to depend on a fortuitous and sometimes unreliable cancellation of error between the lack of consideration of vertex corrections, which tends to cause under-screening in $W_0$, and the systematic underestimation of band gaps computed from semi-local functionals, which tends to cause over-screening \cite{shishkinSelfconsistentGWCalculations2007,kotaniQuasiparticleSelfconsistentGW2007}.

One way to address the issue of starting point dependence is to construct the self-energy in a more self-consistent manner, leading to the development of methods like eigenvalue self-consistent $GW$ (ev$GW$) \cite{hybertsenElectronCorrelationSemiconductors1986,luoQuasiparticleBandStructure2002a,shishkinSelfconsistentGWCalculations2007} and quasiparticle self-consistent $GW$ (QS$GW$) \cite{faleevAllElectronSelfConsistentGW2004,vanschilfgaardeQuasiparticleSelfConsistentGW2006b,kotaniQuasiparticleSelfconsistentGW2007}. In ev$GW$, the eigenvalues used to construct $G$ and $W$ are iterated to self-consistency. Though ev$GW$ is noticeably less dependent on the starting point used, the wave functions used in constructing $G$ and $W$ are not updated in this approach, leading to a residual modest starting point dependence (e.g., $0.4$ eV in the case of azabenzenes \cite{maromBenchmarkGWMethods2012a}). On the other hand, QS$GW$ seeks to variationally minimize the difference between the self-energy and a static nonlocal potential by updating both the wave functions and eigenvalues used to construct $G$ and $W$ and has been shown to be mostly independent of the starting point used \cite{brunevalEffectSelfconsistencyQuasiparticles2006b} (though there do exist questions as to whether this holds true for some metal oxides \cite{liaoTestingVariationsGW2011,isseroffImportanceReferenceHamiltonians2012b}). While iterating on $G$ and/or $W$ provides more consistent results, it also requires greater computational resources. Additionally, while the self-consistent correction of the QP eigenvalues accounts for the error due to DFT band gap underestimation in these methods, it does not systematically account for the lack of vertex corrections, leading to under-screening and larger QP band gaps \cite{shishkinAccurateQuasiparticleSpectra2007,shishkinSelfconsistentGWCalculations2007}. For example, Grumet et al.\ report that ev$GW$ and QS$GW$ overestimate QP gaps by $1.0$ eV and $0.8$ eV on average, respectively \cite{grumetQuasiparticleApproximationFully2018}.

While $GW$ self-consistency schemes can reduce the starting-point dependence of $G_0W_0$, the increased cost of going beyond $G_0W_0$ has incentivized the development of starting points for $G_0W_0$ calculations which do not suffer from the same level of starting-point dependence \cite{rinkeCombiningGWcalculationsExactexchangeDensityfunctional2005b,fuchsQuasiparticleBandStructure2007b,korzdorferStrategyFindingReliable2012,atallaHybridDensityFunctional2013,dauthPiecewiseLinearityGW2016a,knightAccurateIonizationPotentials2016,leppertPredictiveBandGaps2019a,hellgrenElectronicStructureMathrmTiSe2021}. In particular, hybrid DFT functionals, which include exact exchange, are an appealing candidate for improved $G_0W_0$ starting points for multiple reasons.  For example, the GKS band gaps computed with these functionals vary with the amount of exact exchange present, and therefore can be used to remedy the over-screening due to band gap underestimation that is present in semi-local functionals \cite{golzeGWCompendiumPractical2019}. Moreover, hybrid functionals can better address the starting point dependence associated with more localized $d$ states \cite{rinkeCombiningGWcalculationsExactexchangeDensityfunctional2005b,shishkinSelfconsistentGWCalculations2007,jiangFirstprinciplesModelingLocalized2010}, where self-interaction errors present in semi-local functionals are more pronounced \cite{rinkeExcitingProspectsSolids2008a} and lead to spurious orbital energy ordering that can propagate to the $GW$ eigenspectrum \cite{golzeGWCompendiumPractical2019}. In such cases, the presence of exact exchange can help to reduce this error  \cite{rinkeCombiningGWcalculationsExactexchangeDensityfunctional2005b,maromElectronicStructureCopper2011,hellgrenElectronicStructureMathrmTiSe2021,maromBenchmarkGWMethods2012a,korzdorferStrategyFindingReliable2012,luftnerExperimentalTheoreticalElectronic2014a}. 

The use of hybrid functionals like PBE0 \cite{adamoReliableDensityFunctional1999} and HSE \cite{krukauInfluenceExchangeScreening2006} as starting points for $G_0W_0$ calculations has been shown to generally improve agreement with experiment \cite{fuchsQuasiparticleBandStructure2007b,leppertPredictiveBandGaps2019a}.
Moreover, some hybrid functionals can be tuned \cite{steinFundamentalGapsFinite2010} to satisfy the ionization potential (IP) theorem \cite{levyExactDifferentialEquation1984a,almbladhExactResultsCharge1985a}, suggesting the possibility of a more physically accurate and consistent starting point eigensystem. Specifically, Wing et al.\ \cite{wingBandGapsCrystalline2021} developed a procedure for parametrizing a class of screened range-separated hybrid (SRSH) functionals capable of accurately predicting the band gaps of solid state materials without empirical parameters, directly from density functional theory. The parametrization is arrived at by capturing the asymptotic limit of the screened exchange potential and by using an ansatz based on the IP theorem which applies to localized Wannier functions in systems with periodic boundary conditions \cite{maUsingWannierFunctions2016a}. This class of Wannier-localized optimally tuned screened range-separated hybrid (WOT-SRSH) functionals has been recently used to calculate the fundamental band gaps of semiconductors and insulators, leading to excellent agreement with experiment, with an MAE of 0.1 eV \cite{wingBandGapsCrystalline2021}.

For molecules, the use of optimally tuned range-separated hybrid functionals which enforce the IP theorem as a starting point for $G_0W_0$, as suggested in  \cite{refaely-abramsonQuasiparticleSpectraNonempirical2012a}, has been shown to be successful \cite{gallandiLongRangeCorrectedDFT2015a,gallandiAccurateIonizationPotentials2016a,knightAccurateIonizationPotentials2016,rangelEvaluatingGWApproximation2016,boisSizeDependenceNonempiricallyTuned2017,rangelAssessmentLowlyingExcitation2017}. However, as of yet, there has not been an analogous exploration of these non-empirical WOT-SRSH starting points which approximately satisfy the IP theorem for $G_0W_0$ calculations of solid-state systems.

Here, we undertake such an exploration and analyze the performance of single-shot $G_0W_0$@WOT-SRSH calculations.
For a series of 15 semiconductor and insulators,
we construct $G_0$ and $W_0$ using WOT-SRSH
and compute band gaps as well as properties associated with states deeper in the valence band manifold such as valence bandwidths and $d$ band positions. We then compare results with experiments and calculations from other DFT starting points. We also discuss how $G_0W_0$ corrections affect the sensitivity of computed bands gaps to ambiguities in the WOT-SRSH tuning procedure. Overall, our calculations demonstrate that a $G_0W_0$@WOT-SRSH approach provides accurate quasiparticle properties for a broad range of materials, opening the door to predictive single-shot $G_0W_0$ calculations for chemically complex solids.

\section{Theory}
\label{sec:theory}
\subsection{DFT}\label{subsec:DFT_theory}
The starting point for our $GW$ calculations are GKS orbitals $\phi_{n\bm{k}}$ and eigenenergies $\epsilon^0_{n\bm{k}}$, where $n$ is the band index and $\bm{k}$ the wave vector. Here, we primarily focus on the SRSH functional scheme \cite{yanaiNewHybridExchange2004,refaely-abramsonGapRenormalizationMolecular2013a,kronikExcitedStatePropertiesMolecular2016,kronikDielectricScreeningMeets2018}. This class of functionals is formulated by partitioning the exchange portion of the Coulomb potential into
\begin{equation}
\label{eq:rsh}
    \frac{1}{r}=\frac{\alpha+\beta\text{erf}\left(\gamma r\right)}{r}+\frac{1-\left[\alpha+\beta\text{erf}\left(\gamma r\right)\right]}{r}.
\end{equation}
This partition introduces three parameters $\alpha$, $\beta$, and $\gamma$, the physical and computational significance of which is discussed shortly.
When implemented in the hybrid functional, the first term of Eq. (\ref{eq:rsh}) is treated explicitly with Fock exchange, whereas the second term is replaced with an approximate semi-local exchange functional \cite{yanaiNewHybridExchange2004,refaely-abramsonGapRenormalizationMolecular2013a}.
In this framework, $\alpha$ regulates the amount of exact exchange in the short range, $\alpha+\beta$ regulates the amount of exact exchange in the long range, and $\gamma$ is the length scale for the transition between these two limits. The correlation component is treated with the same functional used for the semi-local part of the aforementioned exchange partition.
By specifying the values of $\alpha$, $\alpha+\beta$, and $\gamma$, we can recover various well-known hybrid functionals (Table \ref{tab:SRSH}). For example, if the semi-local exchange is based on the PBE functional \cite{perdewGeneralizedGradientApproximation1996}, then setting $\gamma = 0$ produces a global hybrid functional, PBE$\alpha$ \cite{perdewRationaleMixingExact1996,ernzerhofAssessmentPerdewBurke1999}, and if $\alpha=0.25$, PBE0 is obtained.
For $\gamma=0.106$ $a_0^{-1}$, setting $\alpha+\beta=0$ and $\alpha=0.25$ yields the HSE functional \cite{krukauInfluenceExchangeScreening2006}.

\begin{table}[htbp!]
\begin{centering}
\setlength{\tabcolsep}{0.06 in}
\renewcommand{\arraystretch}{1.3}
\begin{tabular}{|c|ccc|}
\toprule
 & $\bm\alpha$ & $\bm{\alpha+\beta}$ & $\bm{\gamma}$ \textbf{($\bm{a_0^{-1}}$)}\\
 \hline
{PBE0} & $0.25$ & $0.25$ & 0\\
{HSE06} & $0.25$ & 0 & $0.106$\\

{WOT-SRSH} & Varies$^{a}$ & $\varepsilon_{\infty}^{-1}$ & Tuned\\
\botrule 
\end{tabular}
\par\end{centering}
\caption{\label{tab:SRSH}Hybrid functionals in the SRSH formalism\\ a: By default $\alpha$ is set to be $0.25$, but in cases where $\alpha+\beta\sim0.25$ the value of $\alpha$ is increased slightly until the IP ansatz can be satisfied. For more details see \cite{wingBandGapsCrystalline2021} or the discussions in sections \ref{subsec:DFT_theory} and \ref{sec:sensitivity}.}
\end{table}

In this paper, we focus on the novel WOT-SRSH formulation \cite{wingBandGapsCrystalline2021} of the SRSH functional. Here, the choice $\alpha+\beta = \varepsilon^{-1}_\infty$, where $\varepsilon^{-1}_\infty$ is the orientationally averaged electronic contribution to the dielectric constant, enforces the asymptotically correct long-range screening in the Coulomb potential \cite{refaely-abramsonGapRenormalizationMolecular2013a,kronikExcitedStatePropertiesMolecular2016,kronikDielectricScreeningMeets2018}. The range-separation parameter $\gamma$ is non-empirically selected by enforcing an ansatz which extends the IP theorem to the removal of an electron from the highest-energy occupied maximally-localized Wannier function (MLWF) \cite{maUsingWannierFunctions2016a}.
By default, we choose $\alpha=0.25$ because, as seen in global hybrids, setting $\alpha =0.25$ has proven effective for many molecular and solid-state systems \cite{perdewRationaleMixingExact1996,heydEnergyBandGaps2005,heydErratumHybridFunctionals2006,skoneSelfconsistentHybridFunctional2014}.
In cases where setting $\alpha=0.25$ does not yield a unique choice of $\gamma$ via the IP ansatz, as is often the case when $\varepsilon_\infty^{-1}\sim 0.25$, $\alpha$ is increased slightly until an optimal value of $\gamma$ that does not approach zero can be found. The need for a lower bound on the size of $\gamma$ is related to the ``$\gamma$ collapse problem" \cite{dequeirozChargetransferExcitationsLowgap2014,bhandariFundamentalGapsCondensedPhase2018}, where small values of $\gamma$ result in an unphysical effectively PBE$\alpha$ hybrid functional if $\gamma^{-1}$ exceeds the size of the unit cell of the calculation. As seen in Table \ref{tab:DFT_param}, the largest value that $\alpha$ needed to be increased to was $0.35$.
With these constraints, WOT-SRSH functionals are a system-specific but non-empirical class of exchange correlation (xc) functionals that result in a GKS eigensystem that consistently and accurately predicts the QP band gaps of solids, compensating by construction for the derivative discontinuity error present in most density functionals \cite{perdewDensityFunctionalTheoryFractional1982,perdewPhysicalContentExact1983,shamDensityFunctionalTheoryEnergy1983,seidlGeneralizedKohnShamSchemes1996a,onidaElectronicExcitationsDensityfunctional2002d,cohenFractionalChargePerspective2008,kummelOrbitaldependentDensityFunctionals2008b,perdewUnderstandingBandGaps2017}.

\subsection{\texorpdfstring{$GW$}{GW} Method}
In the \textit{ab initio} $GW$ approach, the self-energy $\Sigma=iGW$ of a system is constructed from a DFT GKS eigensystem. As discussed, this GKS eigensystem $\left\{\phi_{n\bm{k}},\epsilon^{\text{DFT}}_{n\bm{k}}\right\}$ depends on the underlying xc functional $V_{xc}$ used to compute it, and by extension the self-energy computed from this eigensystem is also sensitive to the choice of $V_{xc}$. Specifically, the single-particle Green's function $G_0$ is constructed as
\begin{equation}
    G_0(\bm{r},\bm{r}';\omega)=\sum_{n\bm{k}}\frac{\phi_{n\bm{k}}(\bm{r})\phi_{n\bm{k}}^*(\bm{r}')}{\omega-\epsilon_{n\bm{k}}^\text{DFT}\pm i\eta},
\end{equation}
where $\eta$ is a positive infinitesimal real number, and the $\pm$ in front of it is $-$ for occupied states and $+$ for empty states. The dynamically screened Coulomb interaction $W_0$ is given by
\begin{equation}
    W_0(\bm{r},\bm{r}';\omega)=\int d\bm{r}''\varepsilon^{-1}(\bm{r},\bm{r}'';\omega)v(\bm{r}',\bm{r}''),
\end{equation}
where $v(\bm{r},\bm{r}')=\left|\bm{r}-\bm{r}'\right|^{-1}$ and
where the dielectric function,
\begin{equation}
\begin{aligned}
    \varepsilon^{-1}(\bm{r},\bm{r}';\omega)=&\delta(\bm{r},\bm{r}')\\
    &-\int d\bm{r}''v(\bm{r},\bm{r}'')\chi_0(\bm{r}'',\bm{r}',\omega),
\end{aligned}
\end{equation}
is computed within the random-phase approximation (RPA) \cite{hybertsenElectronCorrelationSemiconductors1986} based on the polarizability $\chi_0(\bm{r},\bm{r}',\omega)$, given by the Adler-Wiser expression \cite{adlerQuantumTheoryDielectric1962,wiserDielectricConstantLocal1963}
\begin{widetext}
\begin{equation}
\label{eq:polarizability}
    \chi_0(\bm{r},\bm{r}',\omega)=\sum_{n\bm{k}}^\text{occ.}\sum_{n'\bm{k}'}^\text{emp.}\left[\frac{\phi_{n\bm{k}}^*(\bm{r})\phi_{n'\bm{k}'}(\bm{r})\phi_{n\bm{k}}^*(\bm{r}')\phi_{n'\bm{k}'}(\bm{r}')}{\omega-\left(\epsilon_{n'\bm{k}'}^\text{DFT}-\epsilon_{n\bm{k}}^\text{DFT}\right)+i\eta} -\frac{\phi_{n\bm{k}}(\bm{r})\phi_{n'\bm{k}'}^*(\bm{r})\phi_{n\bm{k}}(\bm{r}')\phi_{n'\bm{k}'}^*(\bm{r}')}{\omega+\left(\epsilon_{n'\bm{k}'}^\text{DFT}-\epsilon_{n\bm{k}}^\text{DFT}\right)-i\eta}\right], 
\end{equation}
\end{widetext}
where the summations are over the occupied and unoccupied bands. In practice, $\chi_0(\bm{r},\bm{r}',\omega)$ is often evaluated statically ($\omega=0$), and a simplified model frequency dependence, such as a plasmon pole model (PPM), is used instead \cite{hybertsenElectronCorrelationSemiconductors1986,godbyMetalinsulatorTransitionKohnSham1989a,oschliesGWSelfenergyCalculations1995}. We also note that a consideration of the denominators in Eq. (\ref{eq:polarizability}) clarifies why the under- or overestimation of the band gap can result in over- or under-screening in $W_0$, respectively.

With the above quantities, the $G_0W_0$ self-energy becomes
\begin{equation}
\begin{aligned}
    \Sigma(\bm{r},\bm{r}';\omega)=\frac{i}{2\pi} \int & d\omega' G_0(\bm{r},\bm{r}';\omega+\omega') W_0(\bm{r},\bm{r}';\omega') \\
    & \times e^{i\omega'\eta} .
\end{aligned}
\end{equation}
This $G_0W_0$ operator can then be used to correct the DFT eigenvalues perturbatively via
\begin{equation}\label{eq:gw_corection}
\epsilon_{n\bm{k}}^\text{QP} = \epsilon^\text{DFT}_{n\bm{k}} + \braket{n\bm{k}|\Sigma(\epsilon_{n\bm{k}}^\text{QP}) - V_{xc}|n\bm{k}}, 
\end{equation}
where, to avoid double counting of beyond-Hartree interactions, the contributions of $V_{xc}$ are subtracted off. Due to the fact that $\epsilon_{n\bm{k}}^\text{QP}$ in Eq. (\ref{eq:gw_corection}) depends on itself, evaluating this expression is non-trivial. However, as is common practice \cite{giantomassiElectronicPropertiesInterfaces2011,liuCubicScalingGW2016,wilhelmGWGaussianPlane2016}, we expand Eq. (\ref{eq:gw_corection}) to first order about $\epsilon^\text{DFT}_{n\bm{k}}$ to evaluate it efficiently.

\section{Computational Details}
\label{sec:calc_details}

\subsection{DFT Calculations}
\label{sec:dft_calcs}
\begin{table}[htbp!]
\begin{centering}
\setlength{\tabcolsep}{0.06 in}
\renewcommand{\arraystretch}{1.25}
\begin{tabular}{|c||ccc|ccc|}
\toprule
 & \multicolumn{3}{c|}{\textbf{\begin{tabular}[c]{@{}c@{}}Lattice\\ Parameters\end{tabular}}} &                                                                                             \multicolumn{3}{c|}{\textbf{\begin{tabular}[c]{@{}c@{}c@{}}WOT-SRSH\\ Parameters$^{\text d}$ \\  \end{tabular}}}                                                 \\ \cline{2-4} \cline{5-7}
 & \textbf{$a$}     & \textbf{$c$}    & \textbf{$u$}                                          & \textbf{$\alpha$} & \textbf{$\beta$} & \textbf{$\gamma$ ($a_0^{-1}$)} \\
 \hline
InSb        & 6.48$^{\text a}$         &              &                                                                                              & 0.25                                     & -0.1745          & 0.17                                                  \\
InAs        & 6.06$^{\text a}$         &              &                                                                                               & 0.25                                     & -0.1623                                 & 0.16                                                  \\
Ge          & 5.66$^{\text a}$         &              &                                                                                              & 0.25                                     & -0.1824                                 & 0.19                                                  \\
GaSb        & 6.1$^{\text a}$          &              &                                                                                              & 0.25                                     & -0.1733                                 & 0.19                                                  \\
Si          & 5.43$^{\text a}$         &              &                                                                                              & 0.25                                     & -0.1611                                 & 0.24                                                  \\
InP         & 5.87$^{\text a}$         &              &                                                                                              & 0.25                                     & -0.1373                                 & 0.23                                                  \\
GaAs        & 5.65$^{\text a}$         &              &                                                                                              & 0.25                                     & -0.1549                                 & 0.15                                                  \\
AlSb        & 6.14$^{\text a}$         &              &                                                                                              & 0.25                                     & -0.1482                                 & 0.14                                                  \\
AlAs        & 5.66$^{\text a}$         &              &                                                                                               & 0.3                                      & -0.1779                                 & 0.18                                                  \\
GaP         & 5.45$^{\text a}$         &              &                                                                                              & 0.25                                     & -0.1375                                 & 0.21                                                  \\
AlP         & 5.47$^{\text a}$         &              &                                                                                              & 0.25                                     & -0.1128                                 & 0.16                                                  \\
C           & 3.57$^{\text a}$         &              &                                                                                              & 0.3                                      & -0.1198                                 & 0.23                                                  \\
AlN         & 3.11$^{\text a}$         & 4.98$^{\text a}$         & 0.3821$^{\text c}$                                             & 0.35                                     & -0.1073                                 & 0.26                                                  \\
MgO         & 4.22$^{\text a}$         &              &                                             & 0.25                                     & $\;$0.0948                                  & 1.5                                                   \\
LiF         & 4.03$^{\text b}$         &              &                                                                     & 0.25                                     & $\;$0.2681                                  & 1.08          \\
\botrule
\end{tabular}
\par\end{centering}
\caption{\label{tab:DFT_param} Parameters used in the DFT starting point calculations. Lattice parameters were taken from experiment, and WOT-SRSH parameters were taken from prior work \cite{wingBandGapsCrystalline2021}.\\
a: \cite{madelungSemiconductorsDataHandbook2004}, b: \cite{reckerDirectionalSolidificationLiFLiBaF31988}, c: \cite{schulzCrystalStructureRefinement1977}, d: \cite{wingBandGapsCrystalline2021}
}
\end{table}
Our DFT calculations are performed using a modified version of the \textsc{QUANTUM ESPRESSO} (version 6.2) plane-wave code \cite{giannozziQUANTUMESPRESSOModular2009b,giannozziAdvancedCapabilitiesMaterials2017c,giannozziQuantumESPRESSOExascale2020} that allows for the use of the SRSH functional \cite{refaely-abramsonGapRenormalizationMolecular2013a} of Eq. (\ref{eq:rsh}) with arbitrary $\alpha$, $\beta$, and $\gamma$ parameters. Other modifications also allow for a more efficient calculation of many hundreds of unoccupied states for GKS systems using adaptively compressed exchange \cite{linAdaptivelyCompressedExchange2016} via what amounts to a non-self-consistent field calculation once the occupied orbitals and ground state density have been converged (see SI Section S-I \cite{supplementaryinformation} for more details). All calculations utilize fully relativistic optimized norm-conserving Vanderbilt pseudopotentials \cite{hamannOptimizedNormconservingVanderbilt2013} obtained from the PSEUDO-DOJO repository \cite{vansettenPseudoDojoTrainingGrading2018}. Using these pseudopotentials, the effects of spin-orbit coupling (SOC) are included self-consistently at the DFT level for all calculated observables. For Ge, Ga, In, Sb, and As, the electrons within a complete set of semi-core shells of the same principal quantum number are treated as valence electrons. For calculations using hybrid functionals and the $GW$ methods, the explicit consideration of these deeper states has been shown to be necessary for the accurate description of the electronic structure of such systems \cite{rohlfingQuasiparticleBandStructure1995,luoQuasiparticleBandStructure2002a,tiagoEffectSemicoreOrbitals2004b,fleszarElectronicStructureII2005}. A plane wave energy cutoff of $135$ Ry and experimental room temperature lattice parameters (summarized in Table \ref{tab:DFT_param}) are used for all systems.

For hybrid functionals, the energy cutoff involved in constructing the exact exchange operator is lowered, without significantly impacting the results at the DFT or $G_0W_0$@DFT levels, from its default value of four times the plane wave energy cutoff to $150$ Ry. In some rare cases where this causes numerical instability in the self-consistent evaluation of the exchange, namely computing the PBE0 starting points for Ge and InAs, this cutoff is raised to the default value of four times the plane wave energy cutoff.

\subsection{\texorpdfstring{$GW$}{GW} Calculations}
\label{subsec:GW_calcs}
All our $GW$ calculations are carried out using the \textsc{BerkeleyGW} package \cite{hybertsenElectronCorrelationSemiconductors1986,deslippeBerkeleyGWMassivelyParallel2012b}. In an effort to minimize the cost of computing many hundreds of unoccupied states using hybrid functionals, the dielectric function is constructed using a symmetry-reduced unshifted Monkhorst-Pack $\bm{q}$ grid. Frequency dependence in the dielectric function is included approximately via the Godby-Needs PPM \cite{godbyMetalinsulatorTransitionKohnSham1989a,oschliesGWSelfenergyCalculations1995}, which has been shown to reproduce the computed band gaps of full-frequency integration at reduced cost \cite{larsonRolePlasmonpoleModel2013a}. It should be noted, however, that this comparable level of accuracy can wane for deeper valence states; previous studies \cite{miglioEffectsPlasmonPole2012,laasnerG0W0bandStructureCdWO42014} report that valence bandwidths and $d$ band binding energies computed using the Godby-Needs PPM are modestly overestimated relative to full-frequency integration.

The static remainder approximation to $\Sigma$ \cite{deslippeCoulombholeSummationsEnergies2013} is used whenever it yields faster convergence with respect to the number of bands, which is the case for all materials except AlN, MgO, and LiF. The band gaps of all materials are converged within (or well within) $50$ meV with respect to the number of bands used to construct $\varepsilon$ and $\Sigma$, the energy cutoff in the construction of $\varepsilon$, and the unshifted $\bm{k}$ grid being used. For more convergence details, see SI Section S-II \cite{supplementaryinformation}.

The effects of SOC are computed at the DFT level and added perturbatively at the $G_0W_0$ level for all materials, an approximation which has precedent and justification for the classes of materials under study \cite{maloneQuasiparticleSemiconductorBand2013a,barkerElectronicOpticalProperties2018,wingComparingTimedependentDensity2019b}. While \textsc{BerkeleyGW} does allow for the explicit computation of SOC effects at the $GW$ level, this would require twice as many bands in the starting point eigensystem, quadrupling the cost of already expensive calculations. However, we find the error of including SOC perturbatively to be minimal. For example we report that for AlSb, a system with a strong SOC band gap renormalization of $240$ meV, the error in the computed SOC renormalization of the band gap is only $6$ meV. For systems with weaker renormalizations like GaP, this error is only $1$ meV.

For band structures with conduction band minima off high symmetry points (as is the case for Si, C, GaP, and AlSb), eigenvalues are interpolated using the \textsc{Wannier90} code \cite{mostofiUpdatedVersionWannier902014}. Due to the similarity in orbital character of the states near the band gap for all the aforementioned systems, only the four highest occupied and four lowest unoccupied bands about the band edges are Wannierized, with $sp^{3}$ starting projections being used for all of them. SOC corrections to the interpolated bands are determined for each eigenenergy $\epsilon_{n\bm{k}}$ and interpolated using MLWFs, as outlined by Malone and Cohen \ \cite{maloneQuasiparticleSemiconductorBand2013a}.

\begin{table*}[p]
\begin{centering}
\def\arraystretch{1.07}
\setlength\tabcolsep{0.06in}
\begin{tabular}{|c||ccc||cc||cc||cc|}
\toprule
		               & \textbf{\begin{tabular}[c]{@{}c@{}}$\bm G_0W_0$@\\      PBE\end{tabular}} & \textbf{\begin{tabular}[c]{@{}c@{}}$\bm G_0W_0$@\\ PBE0\end{tabular}} & \textbf{\begin{tabular}[c]{@{}c@{}}$\bm G_0W_0$@\\ HSE\end{tabular}} & \textbf{WOT-SRSH} & \textbf{\begin{tabular}[c]{@{}c@{}}$\bm G_0W_0$@\\ WOT-SRSH\end{tabular}} & \textbf{ev$GW$} & \textbf{QS$GW$} & \textbf{Ref} & \textbf{Expt, ZPR} \\ \hline
InSb           & 0.09        & 0.58      & 0.45      & 0.32       & 0.44        & 0.79$^{\text a}$          & 0.61$^{\text a}$          & 0.19          & 0.17$^{\text e}$, 0.02$^{\text j}$         \\
InAs           & 0.13        & 0.68      & 0.50      & 0.42       & 0.48        & ---        			      & 0.66$^{\text c}$          & 0.37          & 0.35$^{\text e}$, 0.02$^{\text j}$          \\
Ge             & 0.47        & 0.91      & 0.78      & 0.69       & 0.74        & 0.95$^{\text b}$          & 0.95$^{\text c}$          & 0.71          & 0.66$^{\text f}$, 0.05$^{\text j}$          \\
GaSb           & 0.46        & 1.00      & 0.88      & 0.69       & 0.86        & ---	    			   	 & 1.15$^{\text c}$           & 0.76          & 0.73$^{\text e}$, 0.03$^{\text j}$          \\
Si             & 1.18        & 1.57      & 1.42      & 1.14       & 1.40        & 2.18$^{\text a}$          & 1.49$^{\text a}$          & 1.18          & 1.12$^{\text f}$, 0.06$^{\text j}$          \\
InP            & 1.41        & 1.96      & 1.81      & 1.56       & 1.80        & 1.97$^{\text a}$          & 1.64$^{\text a}$          & 1.40          & 1.35$^{\text e}$, 0.05$^{\text j}$          \\
GaAs           & 1.01        & 1.59      & 1.46      & 1.41       & 1.48        & 1.85$^{\text b}$          & 1.96$^{\text c}$          & 1.47          & 1.42$^{\text e}$, 0.05$^{\text j}$          \\
AlSb           & 1.51        & 1.90      & 1.74      & 1.71       & 1.78        & 2.61$^{\text a}$          & 2.22$^{\text a}$          & 1.65          & 1.61$^{\text e}$, 0.04$^{\text j}$          \\
AlAs           & 2.04        & 2.49      & 2.33      & 2.25       & 2.41        & 2.98$^{\text a}$          & 2.84$^{\text a}$          & 2.20          & 2.16$^{\text e}$, 0.04$^{\text j}$          \\
GaP            & 2.34        & 2.75      & 2.60      & 2.39       & 2.61        & 2.77$^{\text a}$          & 2.67$^{\text a}$          & 2.35          & 2.27$^{\text e}$, 0.08$^{\text j}$          \\
AlP            & 2.44        & 2.92      & 2.75      & 2.52       & 2.82        & 3.2$^{\text a}$           & 2.94$^{\text a}$          & 2.51          & 2.49$^{\text e}$, 0.02$^{\text j}$          \\
C              & 5.58        & 5.95      & 5.82      & 5.76       & 5.92        & 6.41$^{\text a}$          & 6.43$^{\text a}$          & 5.85          & 5.47$^{\text g}$, 0.38$^{\text k}$          \\
AlN            & 5.72        & 6.55      & 6.35      & 6.56       & 6.69        & ---          		 		& 6.80$^{\text c}$        & 6.52          & 6.14$^{\text e}$, 0.38$^{\text k}$         \\
MgO            & 6.96        & 8.07      & 7.99      & 8.16       & 8.62        & 9.53$^{\text a}$          & 9.58$^{\text a}$          & 8.36          & 7.83$^{\text h}$, 0.53$^{\text l}$         \\
LiF            & 13.58       & 14.75     & 14.55     & 15.34      & 15.63       & 15.90$^{\text b}$         & 16.63$^{\text d}$         & 15.35         & 14.20$^{\text i}$, 1.15$^{\text l}$        \\
\hline                                               
MAE            & 0.40        & 0.31      & 0.22      & 0.07       & 0.19        & 0.66$^{\text{m}}$                  & 0.51                      &               & 												\\
MSE 	 	   & -0.40 		 & 0.19 	 & 0.038     & 0.003 	   & 0.19  		 & 0.66$^{\text{m}}$ 				 & 0.51 					 & 				 & 												 \\
Max Error      & -1.77       & -0.60     & -0.80     & -0.20      & 0.40        & 1.17$^{\text{m}}$          		 & 1.28 					 & 				 &  											\\
\botrule
\end{tabular}
\caption{\label{tab:qp_gaps} QP band gaps (in eV) at the WOT-SRSH and $G_0W_0$@DFT level for the various compounds and functionals under study. At the bottom of the table are the MAE (mean absolute error), MSE (mean signed error), and Max Error; all are in eV and measured relative to the reported reference values, which are arrived at by incorporating ZPR corrections into experimental band gap data. Experimental results are arrived at via an analysis of optical absorption spectroscopy data, where excitonic effects are taken into account to arrive at the fundamental gap (see \cite{wingBandGapsCrystalline2021} for details).\\
a: \cite{grumetQuasiparticleApproximationFully2018}, b: \cite{shishkinAccurateQuasiparticleSpectra2007}, c: \cite{vanschilfgaardeQuasiparticleSelfConsistentGW2006b}, d: \cite{kutepovSelfconsistentSolutionHedin2017b}, e: \cite{vurgaftmanBandParametersIII2001a}, f: \cite{madelungSemiconductorsDataHandbook2004}, g: \cite{clarkIntrinsicEdgeAbsorption1964}, h: \cite{whitedExcitonThermoreflectanceMgO1973a}, i: \cite{piacentiniNewInterpretationFundamental1975}, j: \cite{cardonaIsotopeEffectsOptical2005a}, k: \cite{ponceTemperatureDependenceElectronic2015}, l: \cite{chenNonempiricalDielectricdependentHybrid2018b,neryQuasiparticlesPhononSatellites2018a}\\
m: The MAE, MSE, and Max Error for ev$GW$ were computed using the available data for 12 out of 15 compounds.
}
\end{centering}
\end{table*}

\begin{figure*}[p]
\includegraphics[scale=0.295]{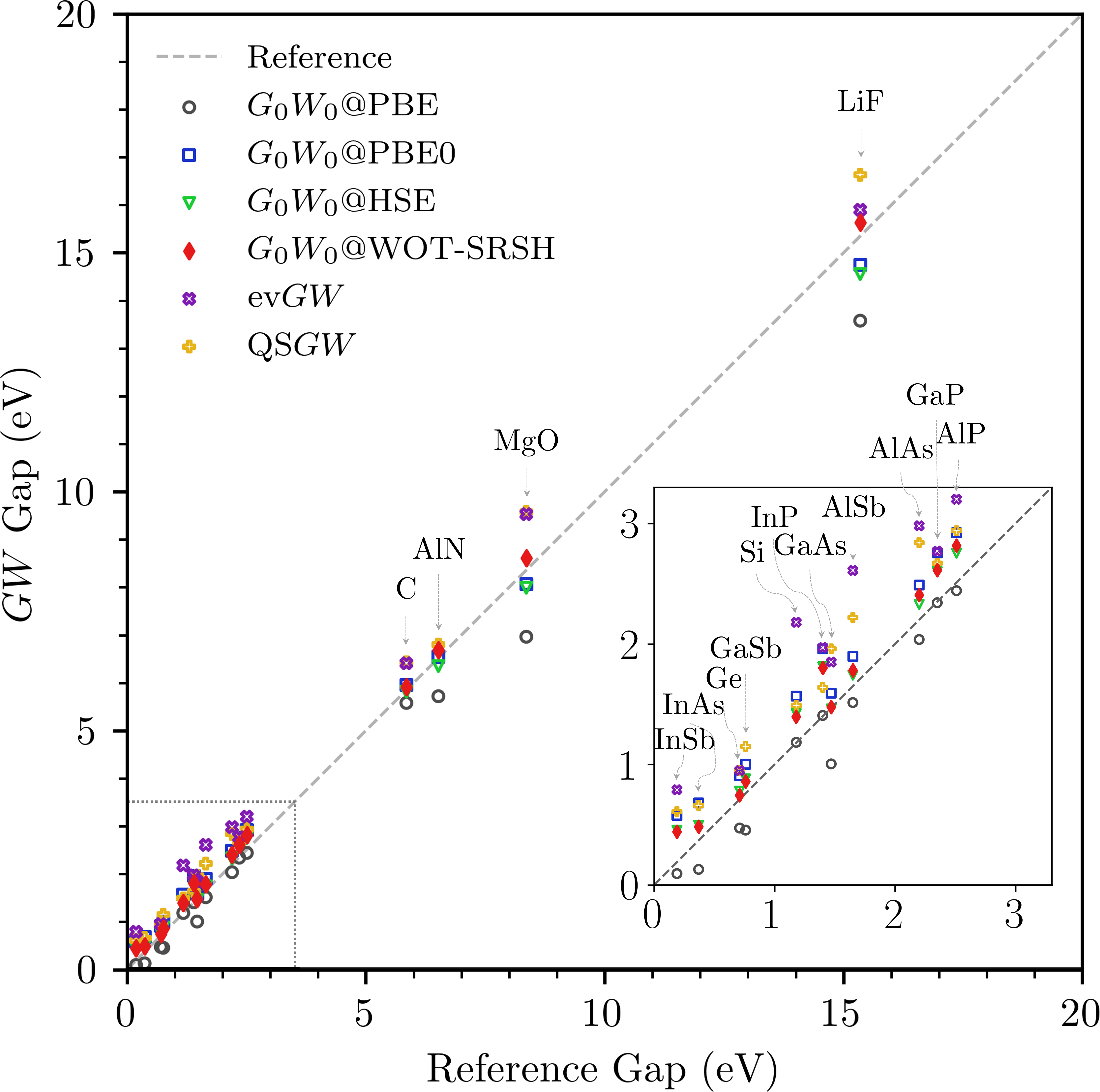}
\caption{\label{fig:scgw_comp} QP band gaps computed using $G_0W_0$@PBE, $G_0W_0$@PBE0, $G_0W_0$@HSE, $G_0W_0$@WOT-SRSH, ev$GW$, and $QSGW$ in reference to ZPR corrected experimental results. Data are taken from table \ref{tab:qp_gaps}. The inset in the lower right corner is a zoom-in of the bottom 3 eV of the data set.}
\end{figure*}

\section{Results and Discussion}
\subsection{Quasiparticle Band Structures}
\subsubsection{Band Gaps}
The QP band gaps of 15 semiconductors and insulators calculated from WOT-SRSH, as well as $G_0W_0$ using four starting points (PBE, PBE0, HSE06, and WOT-SRSH), are given in Table \ref{tab:qp_gaps} and graphed in Figure \ref{fig:scgw_comp}.  Additional results from DFT functionals other than WOT-SRSH can be found in the SI, section S-III.A \cite{supplementaryinformation}. While zero point renormalization (ZPR) effects due to electron-phonon coupling have a significant impact on the band gaps of many solids \cite{giustinoElectronPhononRenormalizationDirect2010a,cannucciaEffectQuantumZeroPoint2011,bottiStrongRenormalizationElectronic2013,antoniusManyBodyEffectsZeroPoint2014,kawaiElectronelectronElectronphononCorrelation2014}, they are not addressed computationally in this paper. Instead, computed band gaps are compared to reference band gaps which remove ZPR effects from the experimental measurements (see \cite{wingBandGapsCrystalline2021}).  Additionally, excitonic effects are accounted for in our reference set by adding estimated or calculated exciton binding energies to the measured optical absorption edge or by inferring the fundamental band gap position based on the location and identification of excitonic absorption peaks in experimental data (See \cite{wingBandGapsCrystalline2021} for more details).

In line with what we have reported previously \cite{wingBandGapsCrystalline2021}, the WOT-SRSH functional yields an excellent MAE of 0.07 eV and a mean signed error (MSE) of 0.00 eV for band gaps---the highest accuracy of all of the methods under study for this set of solids. As the MSE indicates, the data are nearly equally spread between over- and underestimating band gaps. Also, unlike the other functionals, WOT-SRSH has accuracy that is maintained for wider-band gap systems and has a much smaller maximum magnitude error of 0.2 eV.

Performing $G_0W_0$ based on the WOT-SRSH starting point for this set of materials yields an MAE of 0.19 eV, with the $G_0W_0$@WOT-SRSH calculated band gaps maintaining a similar level of precision with a maximum error of 0.40 eV. Notably, the $G_0W_0$@WOT-SRSH band gaps are all slightly overestimated, consistent with the overestimation observed with more rigorously self-consistent methods such as ev$GW$ and QS$GW$ \cite{grumetQuasiparticleApproximationFully2018,shishkinAccurateQuasiparticleSpectra2007,vanschilfgaardeQuasiparticleSelfConsistentGW2006b,kutepovSelfconsistentSolutionHedin2017b} (see table \ref{tab:qp_gaps} and Figure \ref{fig:scgw_comp}). Some of the reported overestimation for these methods has been attributed to the absence of ZPR effects in the band gap \cite{vanschilfgaardeQuasiparticleSelfConsistentGW2006b,shishkinAccurateQuasiparticleSpectra2007}, but our reference band gap accounts for ZPR effects and still indicates some systematic overestimation. However, it is also known that the RPA dielectric function can under-screen and thus overestimate band gaps. As previously noted \cite{shishkinAccurateQuasiparticleSpectra2007, kutepovElectronicStructureNa2016, kutepovSelfconsistentSolutionHedin2017b, maggioGWVertexCorrected2017,schmidtSimpleVertexCorrection2017, kutepovFullQuasiparticleSelfconsistency2022}, beyond-RPA vertex corrections for a similar set of semiconductors and insulators can provide an improvement in the accuracy of the screening and QP band gaps once a consistent starting point that no longer underestimates the band gap is reached.

Comparing $G_0W_0$@WOT-SRSH to self-consistent $GW$ approaches in Figure \ref{fig:scgw_comp}, we find excellent agreement and superior performance relative to experiment for the systems studied here, at a lower computational cost. $G_0W_0$@WOT-SRSH also has a similar qualitative performance to these methods, consistently modestly overestimating band gaps across a broad range of materials.

As is well known and in agreement with prior work \cite{fuchsQuasiparticleBandStructure2007b,chenAccurateBandGaps2015a, jiangGWLinearizedAugmented2016,grumetQuasiparticleApproximationFully2018,golzeGWCompendiumPractical2019}, $G_0W_0$@PBE significantly improves the accuracy of PBE band gaps, in this case bringing its MAE from 1.5 to 0.4 eV. $G_0W_0$@PBE also corrects major qualitative issues such as the inverted band gaps of InSb, InAs, and GaSb. Notably, however, band gaps of some insulators are still underestimated by more than 1 eV (e.g. MgO, LiF) by $G_0W_0$@PBE, leading to a substantial max error of -1.77 eV.

\begin{figure}[b!]
\includegraphics[scale=0.31]{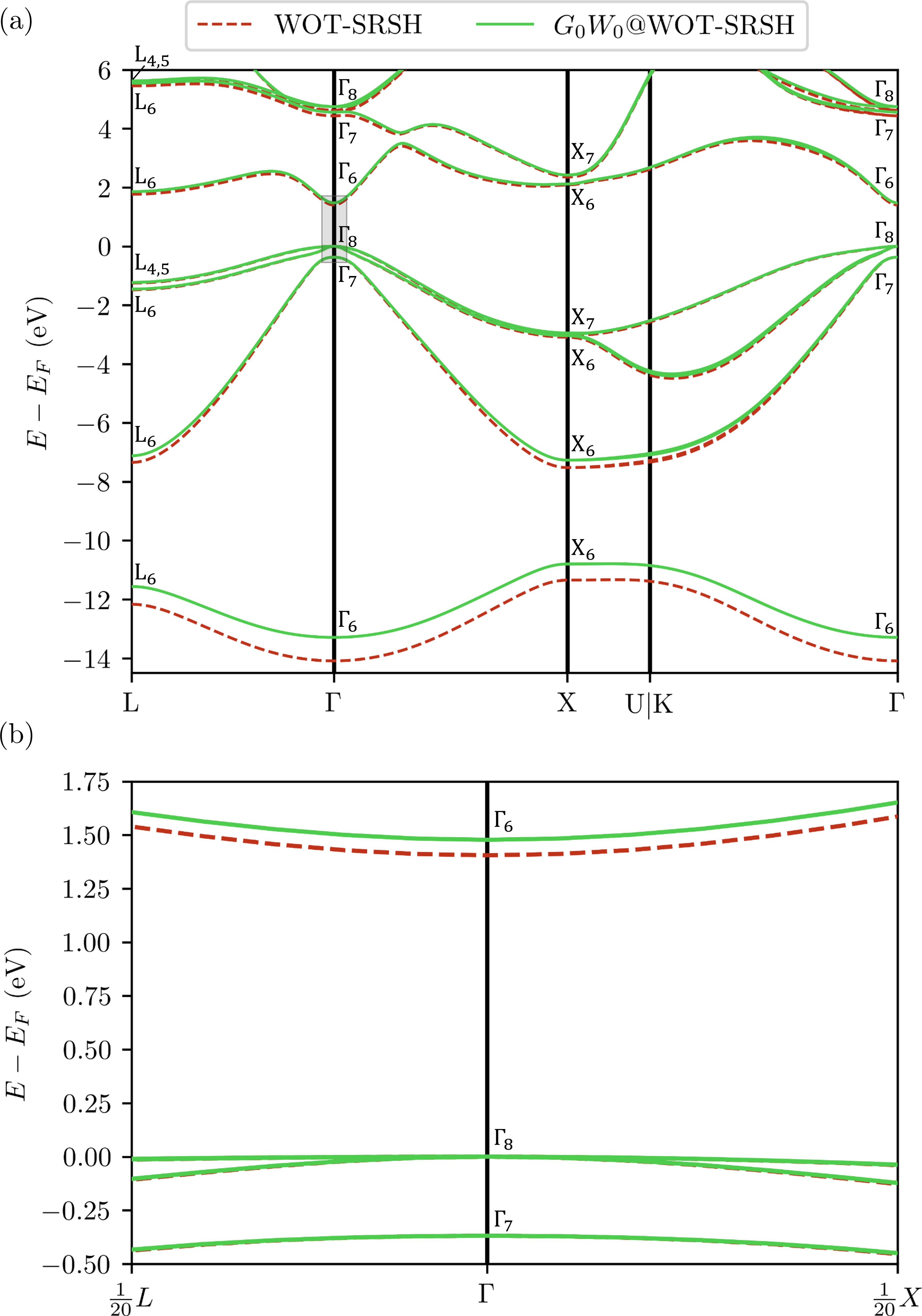}
\caption{\label{fig:bands_plot}Bandstructure of GaAs, including SOC, from WOT-SRSH and $G_0W_0$@WOT-SRSH. a) A full plot of the bandstructure, including the eight highest occupied orbitals and the first few eV of the next eight unoccupied orbitals. b) A zoomed-in inset---depicted by a gray box in a)---of the valence and conduction band extrema. For both plots, $E_F$ is taken to be the energy of the valence band maxima}
\end{figure}

For the insulators studied (C, AlN, MgO, and LiF) we find that well-established hybrid functionals such as HSE and PBE0 offer a significant improvement over PBE as starting points for $G_0W_0$ calculations. However, these hybrids perform slightly worse than PBE for the remaining semiconductors. Overall, for $G_0W_0$@PBE0, we observe a MAE of 0.31 eV, an MSE of 0.19 eV, and a max error of -0.60 eV. The smaller MSE relative to the MAE reflects that the band gaps computed from $G_0W_0$@PBE0 switch from being overestimated for small-band-gap semiconductors to being underestimated for wide-band-gap insulators.

In the case of $G_0W_0$@HSE, calculated QP band gaps have an MAE of 0.22 eV and an MSE of 0.04 eV. This comparatively near-zero MSE reflects that the band gaps computed from $G_0W_0$@HSE switch from being overestimated for small-band-gap semiconductors to being underestimated for wide-band-gap insulators. Moreover, while the MAE calculated for $G_0W_0$@HSE is quite low and comparable to what is seen for $G_0W_0$@WOT-SRSH, the performance of $G_0W_0$@HSE is not consistent. For wide-band-gap insulators such as LiF, $G_0W_0$@HSE underestimates the band gap by nearly 1 eV, leading to a max error of -0.80 eV.

\begin{table*}[!p]
\begin{centering}
\def\arraystretch{1.2}
\setlength\tabcolsep{0.06in}
\begin{tabular}{|c||ccc||cc||c|}
\toprule
								    & \textbf{\begin{tabular}[c]{@{}c@{}}$\bm G_0W_0$@\\      PBE\end{tabular}} & \textbf{\begin{tabular}[c]{@{}c@{}}$\bm G_0W_0$@\\ PBE0\end{tabular}} & \textbf{\begin{tabular}[c]{@{}c@{}}$\bm G_0W_0$@\\ HSE\end{tabular}} & \textbf{WOT-SRSH}  & \textbf{\begin{tabular}[c]{@{}c@{}}$\bm G_0W_0$@\\ WOT-SRSH\end{tabular}} &  \textbf{Expt.} \\ \hline
InSb                     & 11.30                                       					             & 11.51   																 & 11.32   																& 11.96                                                       & 11.32                          					                              & 11.7$^{\text a}$ (XPS), 10.8$^{\text a}$ (ARPES)           \\
InAs                     & 11.90                                       					             & 12.42   																 & 12.34   																& 13.08                                                       & 12.34                          					                              & 12.3$^{\text a}$ (XPS)                \\
Ge                       & 12.82                                       					             & 13.52   																 & 13.30   																& 14.04                                                       & 13.26                          					                              & 12.6$^{\text a}$ (XPS)                  \\
GaSb                     & 11.74                                       					             & 12.36   																 & 12.16   																& 12.83                                                       & 12.13                          					                              & 11.6$^{\text a}$ (XPS), 11.64$^{\text a}$ (ARPES)                  \\
Si                       & 11.51                                       					             & 12.32   																 & 12.10   																& 13.07                                                       & 12.04                          					                              & 12.5$^{\text a}$  (XPS)                 \\
InP                      & 11.22                                       					             & 11.93   																 & 11.72   																& 12.60                                                       & 11.72                          					                              & 11.0$^{\text a}$ (XPS), 11.4$^{\text a}$ (IPES)              \\
GaAs                     & 12.77                                       					             & 13.45   																 & 13.26   																& 14.09                                                       & 13.29                          					                              & 13.8$^{\text a}$ (XPS), 13.1$^{\text a}$ (ARPES)              \\
AlSb                     & 10.67                                       					             & 11.36   																 & 11.14   																& 12.06                                                       & 11.20                          					                              & \textbf{---}             \\
AlAs                     & 11.64                                       					             & 12.39   																 & 12.17   																& 13.35                                                       & 12.29                          					                              & \textbf{---}              \\
GaP                      & 12.18                                       					             & 12.92   																 & 12.70   																& 13.70                                                       & 12.71                          					                              & 12.5$^{\text a}$ (ARPES)                  \\
AlP                      & 11.00                                       					             & 11.83   																 & 11.59   																& 12.75                                                       & 11.67                          					                              & \textbf{---}               \\
C                        & 22.23                                       					             & 23.23   																 & 23.04   																& 24.02                                                       & 23.25                          					                              & 21$^{\text a}$ (XPS)            \\
AlN                      & 6.55                                        					             & 6.73    																 & 6.69    																& 6.65                                                        & 6.75                           					                              & \textbf{---}               \\
MgO                       & 5.09                                        					             & 5.19    																 & 5.18    																& 5.07                                                        & 5.26                           					                              & 6.5$^{\text b}$ (XPS), 7$^{\text b}$ (XES)                   \\
LiF                       & 3.50                                        					             & 3.50    																 & 3.50    																& 3.30                                                        & 3.51                           					                              & 3.4$^{\text c}$ (XPS)              \\
\hline                                                                                                                                                                                              MAE  							   & 0.59                                                   				     & 0.68																     & 0.64   														      & 1.08  												 	       & 0.65                              						                      &  \\                                                                                  

MSE   							   & -0.24                                                  				     & 0.31																     & 0.16 														      & 0.81  												 	       & 0.18                            						                      & \\
\botrule
\end{tabular}
\caption{\label{fig:widths} QP valence bandwidths (in eV), at the WOT-SRSH and $G_0W_0$@DFT level, for the various compounds and functionals under study. For zinc blende materials, the valence bandwidth is defined as the maximal energy difference between the top four (excluding spin degeneracy) valence bands. For the wurtzite and rock salt compounds, the valence bandwidth is defined as the maximal energy difference between the top three valence bands for LiF and MgO and the top six valence bands for AlN. At the bottom of the table are the MAE and MSE; all are in eV and calculated using the leftmost reported experimental values. Experimental data are obtained via XPS, angle-resolved photo-emission spectroscopy (ARPES), and X-ray emission spectroscopy (XES). Due to a lack of quality data on the contributions of ZPR in these results, we do not attempt to correct for such effects in our analysis.\\
a: \cite{goldmannSubvolume1989b}, b: \cite{kowalczykElectronicStructureSrTiO31977}, c: \cite{roCharacterizationLiFUsing1992}
}
\end{centering}
\end{table*}

\begin{table*}[!p]
\begin{centering}
\def\arraystretch{1.2}
\setlength\tabcolsep{0.06in}
\begin{tabular}{|c||ccc||cc||c|}
\toprule
  & \textbf{\begin{tabular}[c]{@{}c@{}}$\bm G_0W_0$@\\      PBE\end{tabular}} & \textbf{\begin{tabular}[c]{@{}c@{}}$\bm G_0W_0$@\\ PBE0\end{tabular}} & \textbf{\begin{tabular}[c]{@{}c@{}}$\bm G_0W_0$@\\ HSE\end{tabular}} & \textbf{WOT-SRSH} & \textbf{\begin{tabular}[c]{@{}c@{}}$\bm G_0W_0$@\\ WOT-SRSH\end{tabular}} &  \textbf{Expt.} \\ \hline
InSb     & 16.18                                                    & 16.74     & 16.57 & 16.24    & 16.55                                                         & 17.1$^{a}$, 16.98$^{b}$, 17.41$^{c}$         \\
InAs     & 15.31                                                    & 16.1      & 16.04 & 15.8     & 16.03                                                         & 16.9$^{a}$, 17.40$^{c}$, 17.38$^{d}$         \\
Ge       & 26.97                                                    & 28.32     & 28.13 & 27.25    & 28.09                                                         & 29.4$^{f}$                \\
GaSb     & 17.12                                                    & 15.52     & 18.15 & 17.52    & 18.11                                                         & 18.8$^{a}$, 18.9$^{g}$          \\
InP      & 14.86                                                    & 15.71     & 15.57 & 15.37    & 15.55                                                         & 17.1$^{a}$                \\
GaAs     & 16.81                                                    & 17.98     & 17.81 & 17.14    & 17.83                                                         & 18.7$^{a}$, 18.7$^{b}$, 18.82$^{c}$   \\
AlSb     & 29.68                                                    & 30.66     & 30.5  & 30.05    & 30.55                                                         & 31.15$^{e}$, 31.60$^{d}$  \\
AlAs     & 36.66                                                    & 37.97     & 37.82 & 37.56    & 38.03                                                         & 39$^{h}$        \\
GaP      & 16.03                                                    & 17.2      & 17.03 & 16.64    & 17.02                                                         & 18.6$^{a}$, 18.7$^{c}$                \\
\hline
MAE     & 1.92        & 1.19      & 1.03   & 1.48    & 1.01       &                     \\
MSE     & -1.92       & -1.19     & -1.03  & -1.48   & -1.01      &                  \\
\botrule
\end{tabular}
\caption{\label{tab:d_orbital} QP highest $d$ band positions, at the WOT-SRSH and $G_0W_0$@DFT level, for the various functionals and $d$-electron containing compounds under study. At the bottom of the table are the MAE and MSE; all are in eV and measured relative to the leftmost reported experimental values. All experimental data are obtained via X-ray photo-emission spectroscopy (XPS). Due to a lack of quality data on the contributions of ZPR in these results, we do not attempt to correct for such effects in our analysis.\\
a: \cite{shevchikDensitiesValenceStates1974}, b: \cite{cardonaPhotoemissionGaAsInSb1972}, c: \cite{leyTotalValencebandDensities1974}, d: \cite{waldropEffectInterfaceComposition1992}, e: \cite{ehlersAngleresolvedphotoemissionStudyAlSb1989}, f: \cite{krautSemiconductorCorelevelValenceband1983}, g: \cite{gualtieriRayPhotoemissionCore1986}, h: \cite{okumuraPhotoemissionStudiesAlAs1987}
}
\end{centering}
\end{table*}

\subsubsection{Band Structure} 
\label{subsubsec: bands}
In Figure \ref{fig:bands_plot} we plot the calculated band structures from WOT-SRSH and $G_0W_0$@WOT-SRSH for GaAs. Apart from a small shift, the bands are nearly identical. Additionally, the similarity of their curvature, especially near the band gap can be seen in the lower inset plot. $G_0W_0$@WOT-SRSH corrections do, however, result in a flattening of the valence bands compared to those of WOT-SRSH. This can be seen in the top figure, where the lowest valence band from WOT-SRSH is $\sim0.5$ eV lower than its $G_0W_0$@WOT-SRSH counterpart. This indicates that away from the band gap, there may be more significant differences between bandstructures of $G_0W_0$@WOT-SRSH and WOT-SRSH. In sections \ref{subsubsec:bandwidths} and \ref{subsubsec:d-bands} we analyze these differences in greater detail.

\subsubsection{Bandwidths}\label{subsubsec:bandwidths}
The calculated valence bandwidths for all compounds
are reported in Table \ref{fig:widths}. Additional results from DFT functionals other than WOT-SRSH can be found in the SI, section S-III.B \cite{supplementaryinformation}.
For zinc blende materials, where there is strong $sp^3$ hybridization, the valence bandwidth is defined as the maximal energy difference between the top four (excluding spin degeneracy) valence bands. For the wurtzite and rock salt compounds, the valence bandwidth is defined as the maximal energy difference between the top three valence bands for LiF and MgO and the top six valence bands for AlN since it has twice as many atoms per unit cell.
For more information on the states under consideration to compute bandwidths, see the leftmost column in Table \ref{fig:widths}. Unlike for QP band gaps, the effects of ZPR are not incorporated when comparing to experiment. Details on the DFT calculations (excluding WOT-SRSH) can be found in the SI. For WOT-SRSH, the MAE and MSE are 1.08 and 0.81 eV respectively, suggesting the method tends to overestimate valence bandwidths by $\sim 1$ eV.
$G_0W_0$@WOT-SRSH has an MAE of 0.65 eV and an MSE of 0.18 eV, showing that $G_0W_0$ corrections away from the band gap offer a significant improvement in accuracy relative to WOT-SRSH.
Notably, the valence bandwidths for the zinc blende compounds are generally overestimated relative to experiment by both WOT-SRSH and $G_0W_0$@WOT-SRSH, while for the rock salt compounds studied, the valence bandwidths are, if anything, underestimated.

Moving to the well-established starting point functionals, $G_0W_0$@PBE computes bandwidths quite well, with an MAE of 0.59 eV and an MSE of -0.24 eV. It also tends to underestimate bandwidths as its MSE suggests.
For hybrids, $G_0W_0$@PBE0 and $G_0W_0$@HSE have MAEs of 0.68 and 0.64 eV and MSEs of 0.31 and 0.16 eV respectively. Interestingly, $G_0W_0$@HSE and $G_0W_0$@WOT-SRSH have comparable levels of accuracy for bandwidths. Unlike in the case of band gaps, this similar level of accuracy persists for wide-gap insulators.

\subsubsection{\texorpdfstring{$d$}{d} Band Energies}\label{subsubsec:d-bands}
For each semiconductor in our set that has elements for which $d$ orbitals are explicitly treated as valence states, the $d$ band position, defined as the highest $d$ orbital eigen-energies relative to the valence band maxima, is reported in Table \ref{tab:d_orbital}. Additional results from DFT functionals other than WOT-SRSH can be found in the SI, section S-III.C \cite{supplementaryinformation}. As in the case of bandwidths, the effects of ZPR are not incorporated when comparing to experiment. For all calculations, we observe a universal underestimation of the $d$-orbital locations, making the distinction between the MAE and MSE meaningless. $G_0W_0$ corrections offer an improvement in accuracy for all starting points. For WOT-SRSH, the MSE decreases from 1.48 to 1.01 eV. For $G_0W_0$@PBE, it plummets from 3.8 to 1.92 eV. For $G_0W_0$@PBE0 it decreases from 1.7 to 1.19 eV, and for $G_0W_0$@HSE it decreases from 1.47 to 1.03 eV. In total, $G_0W_0$@HSE and $G_0W_0$@WOT-SRSH appear to perform the best and offer a comparable level of accuracy. However, both methods  still deviate from experimental reports by $\sim 1$ eV.

\subsection{Parameter Sensitivity of WOT-SRSH and \texorpdfstring{$G_0W_0$}{G0W0}@WOT-SRSH}
\label{sec:sensitivity}
The IP ansatz used to tune the range-separation parameter in the WOT-SRSH functional determines $\gamma$ uniquely for a given choice of $\alpha$ and $\varepsilon_\infty$. However, there can be ambiguities in the selection of $\alpha$ and $\varepsilon_\infty$, with consequences for the predictive power of WOT-SRSH band gaps. Assuming first that $\varepsilon_\infty$ has been computed accurately and that $\beta$ is set to enforce $\alpha+\beta=\varepsilon_\infty^{-1}$, there exists, in principle, a range of choices of $\alpha$ for each material where one can find an optimal $\gamma>0$ satisfying the IP ansatz. These optimal $(\alpha,\gamma)$ pairs produce band gaps which can differ by up to a few hundred meV. Some of the ambiguity in selecting $\alpha$ is avoided by setting $\alpha=0.25$ by default, but as discussed in Sec. \ref{subsec:DFT_theory} an optimal $\gamma$ cannot always be found when $\alpha=0.25$, especially if $\varepsilon_\infty^{-1}\sim0.25$. In such cases, $\alpha$ must be varied until it becomes possible to find an optimal $\gamma$ which satisfies the IP ansatz and the constraint $\gamma>L^{-1}$, where $L$ is the unit cell size used in the calculations. Additionally, it should be noted that while in principle $\alpha$ can be increased to be as large as $1$, in practice values approaching unity are generally considered to be unphysically large for most systems \cite{wingBandGapsCrystalline2021}. Thus, WOT-SRSH predictions are, in practice, more precise than those one would obtain from considering the full range of $\alpha$ values.

\begin{figure}[t]
\begin{centering}
\includegraphics[scale=0.70]{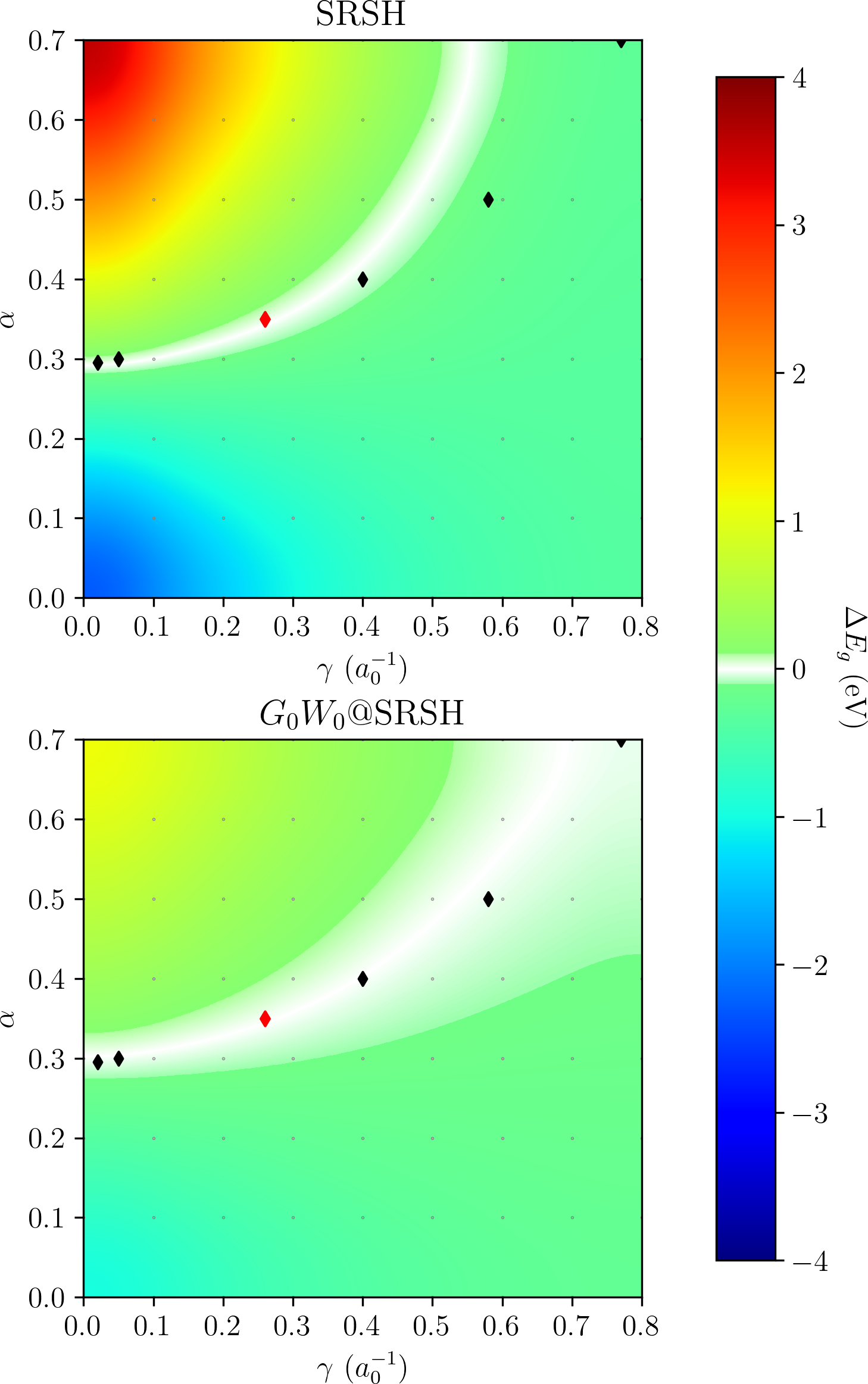}
\par\end{centering}
\caption{\label{fig:alpha_gamma_var}The direct band gap of AlN, relative to a reference value ($\Delta E_g=E_g-E_{g,\text{ref}}$) at the SRSH and $G_0W_0$@SRSH levels, interpolated over a wide range of $(\gamma,\alpha)$ values. The grid of performed calculations is represented as gray dots, and the pairs satisfying the WOT-SRSH constraint are depicted as diamonds, with the reference pair for $\Delta E_g$ in red. A range of $\pm100$ meV about the reference is shaded in white. $G_0W_0$ can be seen to suppress the overall variation at the SRSH level by about a factor of 3. Moreover, the pairs satisfying the WOT-SRSH constraint can be seen to leave the $\pm100$ meV range for the somewhat unphysically large values of $\alpha$ for SRSH, but not for $G_0W_0$@SRSH.}
\end{figure}

Nonetheless, it is of significant interest to explore the ambiguity in selecting $\alpha$ in the WOT-SRSH framework further and its consequences for $G_0W_0$@WOT-SRSH. To do so, we systematically vary $\alpha$ and $\gamma$ and compute GKS and $G_0W_0$ QP band gaps for AlN. AlN is a good candidate for investigation since it has a dielectric constant that is very close to $0.25$ and its band gap exhibits significant variation, on the order of hundreds of meV, between optimal $(\alpha,\gamma)$ pairs. The difference $\Delta E_g=E_g-E_{g,\text{ref}}$ between computed band gaps, relative to the chosen reference gaps $E_{g,\text{ref}}$ for SRSH and $G_0W_0$@SRSH calculations of AlN, can be seen over a range of $\alpha$ and $\gamma$ in Figure \ref{fig:alpha_gamma_var}. Note that, as indicated, we are using, strictly speaking, the SRSH functional, as opposed to WOT-SRSH, meaning the IP ansatz is not satisfied for most of the data shown in Figure \ref{fig:alpha_gamma_var}. The only overall constraint applied here is $\beta=\varepsilon_\infty^{-1}-\alpha$. Additionally, the $G_0W_0$ calculations presented here are slightly under-converged, using 256 bands to construct $\varepsilon$ and $\Sigma$. Pairs of $(\alpha,\gamma)$ satisfying the IP ansatz are marked with black diamonds, and the reference band gap $E_{g,\text{ref}}$ is chosen to be the band gap obtained with the WOT-SRSH parameters of prior work \cite{wingBandGapsCrystalline2021} at either the DFT or $G_0W_0$ level. A range of $\pm100$ meV about this reference value is specified in white in the colormap.

Overall, $G_0W_0$ corrections to the SRSH starting point substantially reduce the sensitivity of the computed band gap to variations in $\alpha$ and $\gamma$ by about a factor of 3. Specifically, at the SRSH level $\Delta E_g$ varies by 6.0 eV for the large ranges of $\alpha$ and $\gamma$ considered, while for $G_0W_0$@SRSH it varies by only 2.14 eV. This reduction in sensitivity becomes much more pronounced when only $(\alpha,\gamma)$ pairs satisfying the IP ansatz are considered. At the WOT-SRSH level, the $\Delta E_g$ values produced by these pairs have a range of $322$ meV and depart from the white $\pm100$ meV range about $E_{g,\text{ref}}$ for the somewhat unphysical larger choices of $\alpha$. In contrast, at the $G_0W_0$@WOT-SRSH level, the exhibited range is only $26$ meV. This reduction is by more than an order of magnitude, and substantially lower than the reduction observed for the overall SRSH functional. A similar set of trends is also observed for the other materials; see SI Section S-IV.1 \cite{supplementaryinformation}. 

We also note that there exists some uncertainty as to how to select the long-range screening $\alpha+\beta=\varepsilon_\infty^{-1}$. For example, one could compute $\varepsilon_\infty$ by considering the head of the RPA dielectric matrix used in $GW$ \cite{hybertsenFirstPrinciplesTheoryQuasiparticles1985}. However, the response to a finite electric field can also be employed, leading to a beyond-RPA value of $\varepsilon_\infty$ \cite{nunesBerryphaseTreatmentHomogeneous2001,souzaFirstPrinciplesApproachInsulators2002}. The inclusion of local field effects for such methods can also significantly affect the calculated response \cite{northrupTheoryQuasiparticleEnergies1987}. Additionally, the underlying DFT functional affects the computed $\varepsilon_\infty$. In fact, it is even possible to self-consistently update the value of $\alpha+\beta$ in an SRSH functional so that it matches the value obtained from a calculation of the dielectric constant using said updated  functional \cite{skoneSelfconsistentHybridFunctional2014, skoneNonempiricalRangeseparatedHybrid2016a}.

Accordingly, we consider the effects of varying the dielectric constant for AlN. For the WOT-SRSH functional used here, $\varepsilon_\infty$ was computed in prior work \cite{wingBandGapsCrystalline2021} via finite electric fields using a PBE0 functional. In lieu of re-computing the optimal $\gamma$ value for different values of $\varepsilon_\infty$, we simply consider the effects of changing $\varepsilon_\infty$ by $\pm 10\%$ while keeping $\alpha$ and $\gamma$ fixed. This choice likely exacerbates the sensitivity of the functional because the IP ansatz is slightly violated for the values of $\varepsilon_\infty$ which differ from the original one used to tune $\gamma$. As can be seen in Section S-IV.2 of the SI \cite{supplementaryinformation}, these perturbations in $\varepsilon_\infty$ result in the band gap changing by 260 meV at the SRSH level but only 80 meV at the $G_0W_0$@SRSH level. This behavior is in line with the approximately threefold band gap range flattening observed above. A similar trend is also observed for the other materials considered in the SI, Section S-IV.2 \cite{supplementaryinformation}.

\section{Conclusions}
We have shown that a new WOT-SRSH class of range-separated hybrid functionals, which is tuned to satisfy an IP ansatz for localized orbitals and to have the correct asymptotic boundary conditions of the Coulomb potential in solids, offers an excellent starting point for $G_0W_0$ calculations of the electronic structure for a wide range of semiconductors and insulators. $G_0W_0$@WOT-SRSH calculations perform at a level of accuracy that is equal to or better than state-of-the-art ev$GW$ and QS$GW$ calculations at a lower computational cost.
Additionally, $G_0W_0$@WOT-SRSH tends to improve the description of states deeper in the valence band manifold, leading to more accurate valence bandwidths and $d$ band locations.
Finally, we have shown that $G_0W_0$@WOT-SRSH corrections greatly reduce the sensitivity of computed bands gaps to variations in the underlying WOT-SRSH parameters that can arise from ambiguities in the optimal tuning procedure. In total, these calculations demonstrate that pairing WOT-SRSH with single-shot $G_0W_0$ methods offers a high-accuracy predictive method for calculating QP properties of materials with a wide range of band gaps.

\section*{Acknowledgements}
This work was supported via US-Israel NSF-Binational Science Foundation (BSF) Grant DMR-2015991. Computational resources were provided by the Extreme Science and Engineering Discovery Environment (XSEDE) \cite{xsede} supercomputer Stampede2 at the Texas Advanced Computing Center (TACC) through Grant No. TG-DMR190070, and additional computational resources were provided by the National Energy Research Scientific Computing Center (NERSC), DOE Office of Science User Facilities supported by the Office of Science of the US Department of Energy under Contract DE-AC02-05CH11231. MRF acknowledges support from the UK Engineering and Physical Sciences Research Council (EPSRC), grant no. EP/V010840/1, and LK thanks the Aryeh and Mintzi Katzman Professorial Chair and the Helen and Martin Kimmel Award for Innovative Investigation.
\bibliography{main}

\end{document}


\title{An Optimally-Tuned Starting Point for Single-Shot $GW$ Calculations
of Solids:\emph{ Supplementary Information}}
\author{Stephen E. Gant}
\affiliation{Department of Physics, University of California, Berkeley, Berkeley,
California 94720, USA}
\author{Jonah B. Haber}
\affiliation{Department of Physics, University of California, Berkeley, Berkeley,
California 94720, USA}
\author{Marina R. Filip}
\affiliation{Department of Physics, University of Oxford, Clarendon Laboratory,
Oxford OX1 3PU, United Kingdom}
\author{Francisca Sagredo}
\affiliation{Department of Physics, University of California, Berkeley, Berkeley,
California 94720, USA}
\author{Dahvyd Wing}
\affiliation{Department of Molecular Chemistry and Materials Science, Weizmann
Institute of Science, Rehovoth 76100, Israel}
\author{Guy Ohad}
\affiliation{Department of Molecular Chemistry and Materials Science, Weizmann
Institute of Science, Rehovoth 76100, Israel}
\author{Leeor Kronik}
\affiliation{Department of Molecular Chemistry and Materials Science, Weizmann
Institute of Science, Rehovoth 76100, Israel}
\author{Jeffrey B. Neaton}
\affiliation{Department of Physics, University of California, Berkeley, Berkeley,
California 94720, USA }
\affiliation{Materials Sciences Division, Lawrence Berkeley National Laboratory,
Berkeley, California 94720, USA}
\affiliation{Kavli Energy NanoScience Institute at Berkeley, Berkeley, California
94720, USA}
\maketitle

\section{Non-Self-Consistent Field Calculations for Hybrid Functionals}

One roadblock for using hybrid functional starting points in \texttt{QUANTUM ESPRESSO}
(QE) for MBPT calculations is that by default in QE (version 6.2),
hybrid DFT can only be carried out self-consistently. More specifically,
this involves two nested self-consistent field (SCF) loops. The inner
loop converges the generalized Kohn-Sham (GKS) eigensystem while holding
the exact exchange operator in the Hamiltonian fixed. Then, the outer
loop constructs the exact exchange operator from the GKS system and
checks for self-consistency. Naively, following QE's constraint that
the computation of the exact exchange for $\mathcal{O}\left(10^{3}\right)$
states be done self-consistently requires approximately seven costly
iterations. However, given a converged ground state density $\rho$
and set of occupied orbitals $\left\{ \phi_{i}\right\} _{i\in\text{occ}}$,
only one iteration should be needed. As is well-known, the conventional
DFT Hamiltonian is simply a functional of $\rho$. Moreover, the exact
Fock exchange operator
\begin{equation}
V_{X}\left(\boldsymbol{r},\boldsymbol{r}'\right)=-\sum_{i\in\text{occ}}\frac{1}{\left|\boldsymbol{r}-\boldsymbol{r}'\right|}\phi_{i}^{*}\left(\boldsymbol{r}'\right)\phi_{i}\left(\boldsymbol{r}\right)
\end{equation}
only depends on the occupied orbitals. Because altering the number
of unoccupied states under consideration will not alter the Hamiltonian,
it should be possible to perform a non-self consistent field (NSCF)
calculation using a converged density and set of occupied orbitals
$\left\{ \rho,\left\{ \phi_{i}\right\} _{i\in\text{occ}}\right\} $
to obtain the unoccupied states needed for $GW$.

Another issue obstructing the aforementioned single iteration of exact
exchange is that QE does not construct the exact exchange operator
until a single inner SCF loop has been completed. This makes even
simply re-running a calculation from a well-converged ground state
density and set of occupied orbitals much more costly than it needs
to be. The QE algorithm re-diagonalizes an incomplete Hamiltonian
without exact exchange, altering the well-converged eigensystem, and
necessitating about as many total SCF iterations to re-converge as
one would need if the system had started from a completely unconverged
$\left\{ \rho,\left\{ \phi_{i}\right\} _{i\in\text{occ}}\right\} $.
Both of the above facts make performing a non-SCF (NSCF) calculation
for the unoccupied bands (a common best practice in $GW$) impossible
for hybrid functionals in QE.

To remedy these issues, we implemented a few alterations to our QE
code and workflow. First, we permitted the exact exchange operator
to be constructed before the first inner SCF loop. This means that
a restarted hybrid functional calculation no longer pushes $\left\{ \rho,\left\{ \phi_{i}\right\} _{i\in\text{occ}}\right\} $
away from its converged state. Second, we modified the code to allow
for explicit control of the number of exact exchange iterations performed.
This was needed because QE will always perform at least two exact
exchange iterations so that it can check for self-consistency in the
outer SCF loop. However, if the eigensystem being fed in is already
converged, the second iteration is unnecessary. Lastly, we devised
a workaround which takes a converged ground state eigensystem $\left\{ \rho,\left\{ \phi_{i}\right\} _{i\in\text{occ}}\right\} $
and populates the desired number of unoccupied states in the saved
many-body eigenfunction. These alterations allow for QE to accurately
compute the new extended eigensystem, including $\mathcal{O}\left(10^{3}\right)$
unoccupied states, through a single SCF loop. In practice, we found
this to reduce the cost of computing the unoccupied states for a $GW$
calculation by about a factor of seven (the number of SCF loops usually
required without the above modifications).

It should also be noted that even though conventional computation
of the exact exchange operator does not depend on the unoccupied states,
there is a small dependence observed in our work. When testing our
modifications on Si we found that doubling the number of bands under
consideration resulted in a change of $\sim10$ $\mu$eV in the computed
band gap of Si. This is likely due to the fact that we are using adaptively
compressed exchange (ACE) \citep{linAdaptivelyCompressedExchange2016}
to handle the exact exchange. In this algorithm, the unoccupied states
are formally incorporated in the construction of the compressed exchange
operator. Moreover, the number of unoccupied states, especially those
close to the Fermi level, have been shown to have a minor influence
on the convergence of the exact exchange operator \citep{linAdaptivelyCompressedExchange2016}.
Nonetheless, such a small change is negligible in comparison to the
other sources of error under consideration.

\section{$GW$ Convergence}

The convergence parameters used in the $G_{0}W_{0}$ calculations
for each system can be found in Table \ref{GW_conv}. \ref{subsec:Number-of-Bands}
and \ref{subsec:-grid} provide an illustrative example for GaAs of
the analysis used to determine the final convergence parameters of
our $G_{0}W_{0}$ calculations. Due to the increased cost of hybrid
functionals, convergence data were determined using the PBE functional
\citep{perdewGeneralizedGradientApproximation1996} for calculations.
This example shows that based on the level of convergence of the number
of $q$ points, the dielectric cutoff, and the number of bands used
in constructing $\varepsilon$ and $\Sigma$, the $G_{0}W_{0}$ calculations
for GaAs underestimate the its QP band gap by about $20$ meV.

\begin{table}[h]
\begin{centering}
\begin{tabular}{|c||ccc|} \toprule   & \textbf{\# Bands} & \textbf{\begin{tabular}[c]{@{}c@{}}$\bm{\varepsilon}$ Cutoff\\ (Ry)\end{tabular}} & \textbf{$\bm k$/$\bm q$-grid} \\ \hline InSb                                                                                   & 512            & 50                                                                     & 6x6x6           \\ InAs                                                                                    & 512            & 40                                                                     & 6x6x6           \\ Ge                                                                                      & 512            & 50                                                                     & 8x8x8           \\ GaSb                                                                                    & 512            & 60                                                                     & 8x8x8           \\ Si                                                                                      & 256            & 30                                                                     & 6x6x6           \\ InP                                                                                     & 256            & 40                                                                     & 6x6x6           \\ GaAs                                                                                    & 512            & 70                                                                     & 6x6x6           \\ AlSb                                                                                    & 256            & 40                                                                     & 6x6x6           \\ AlAs                                                                                    & 512            & 40                                                                     & 6x6x6           \\ GaP                                                                                     & 256            & 30                                                                     & 6x6x6           \\ AlP                                                                                     & 256            & 30                                                                     & 6x6x6           \\ C                                                                                       & 256            & 60                                                                     & 6x6x6           \\ AlN                                                                                     & 1024           & 50                                                                     & 7x7x4           \\ MgO                                                                                     & 512            & 60                                                                     & 6x6x6           \\ LiF                                                                                     & 1024           & 80                                                                     & 6x6x6          \\ \bottomrule
\end{tabular}
\par\end{centering}
\caption{\label{GW_conv}$GW$ convergence data for each system under study.
The static remainder approximation for the self-energy is used for
all systems in which it offered faster convergence with respect to
the number of bands (i.e. all materials but AlN, MgO, and LiF). The
number of bands (including both occupied and unoccupied states) used
are chosen to be powers of $2$ in order to maximize band parallelization
efficiency.}

\end{table}
\newpage{}

\subsection{Number of Bands and Dielectric Cutoff\label{subsec:Number-of-Bands}}

\begin{figure}[h]
\begin{centering}
\includegraphics[scale=0.85]{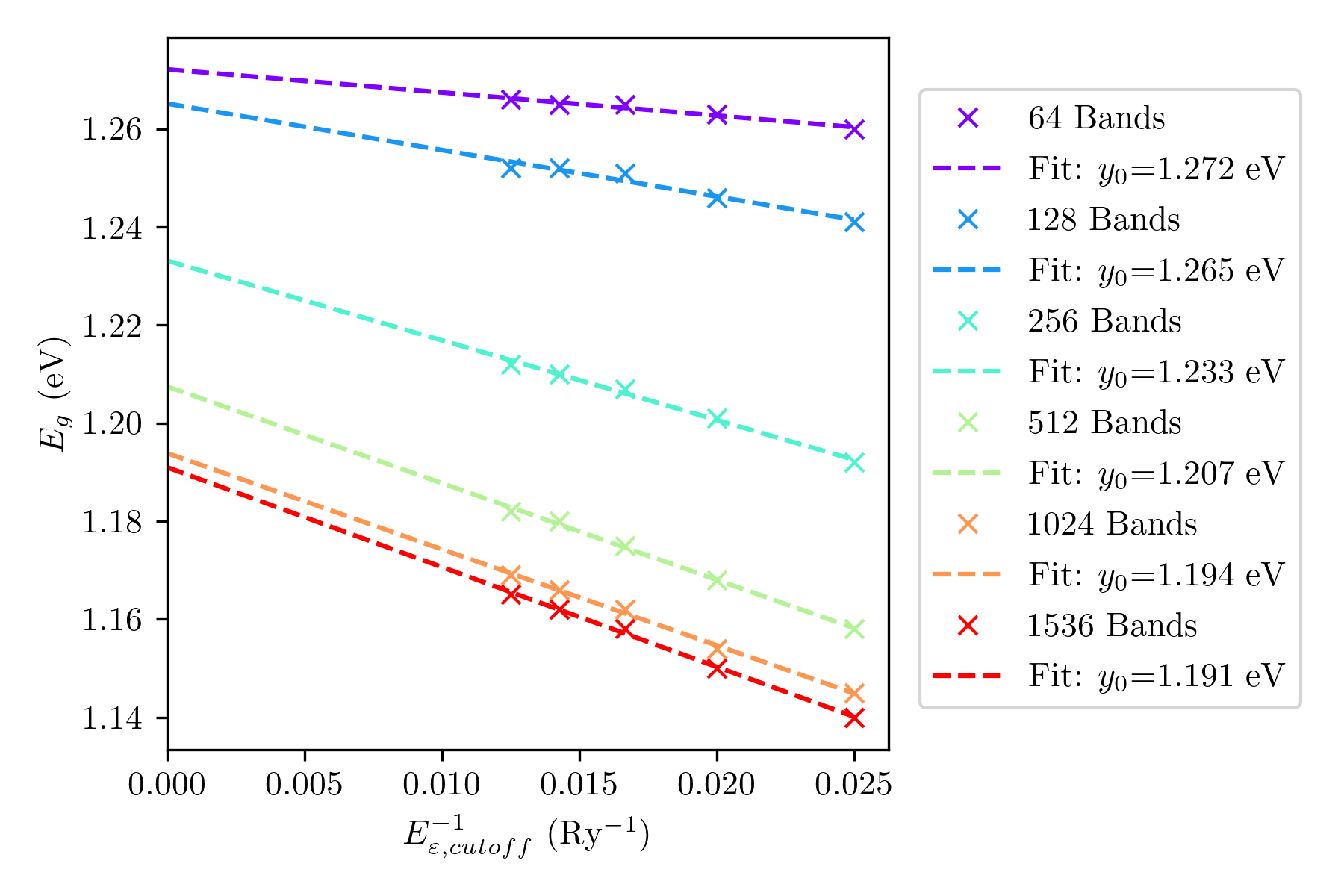}
\par\end{centering}
\caption{\label{fig:Convergence-of-GaAs}Convergence of the direct band gap
of GaAs at $\Gamma$ with respect to the number of bands and the dielectric
cutoff. Dielectric cutoff convergence is extrapolated to infinity
via a linear fit to the inverse cutoff $E_{\varepsilon,\text{cutoff}}^{-1}$.
The $y$ intercept of the fit, representing the $E_{\varepsilon,\text{cutoff}}\rightarrow\infty$
limit, is reported in the legend for varying number of bands.}
\end{figure}

Convergence of the direct band gap of GaAs at $\Gamma$ with respect
to the number of bands (used in both the construction of the polarizability
and the self-energy) and the dielectric cutoff was performed simultaneously.
This yielded dielectric cutoff convergence series for a variety of
numbers of bands, as depicted in Figure \ref{fig:Convergence-of-GaAs}.
In order to extrapolate the limit of $E_{\varepsilon,\text{cutoff}}\rightarrow\infty$
the $y$ intercept of the data series for $E_{\varepsilon,\text{cutoff}}^{-1}$
was considered. Our choice of $E_{\varepsilon,\text{cutoff}}=70$
Ry and $512$ bands for sufficiently converged parameters is expected
to underestimate the fully converged gap by about 10 meV.
\newpage{}

\subsection{$q$-grid\label{subsec:-grid}}

\begin{figure}[h]
\begin{centering}
\includegraphics[scale=0.28]{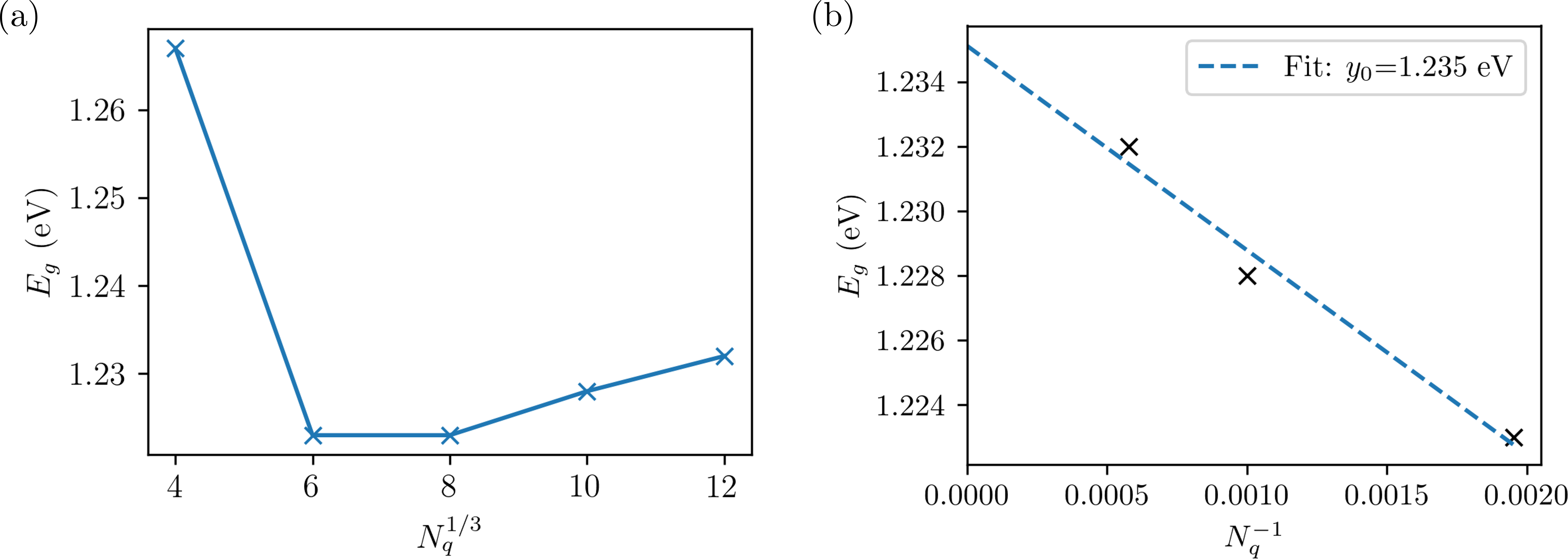}
\par\end{centering}
\caption{\label{fig:kgrid-conv-of-GaAs}Convergence of the direct band gap
of GaAs at $\Gamma$ with respect to the number of $q$-points $N_{q}$
used in constructing the dielectric response function. (a) shows the
relationship with respect to $N_{q}^{\nicefrac{1}{3}}$, and (b) shows
the relationship with respect to $N_{q}^{-1}$.}
\end{figure}

Using a half-shifted 10$\times$10$\times$10 $k$-grid to obtain
a well-converged ground state density at the PBE level, the convergence
of the direct band gap of GaAs at $\Gamma$ with respect to the number
of $q$ points used in constructing the dielectric response function
was analyzed. Figure \ref{fig:kgrid-conv-of-GaAs} (a) shows that
while the band gap is still changing even at $N_{q}^{\nicefrac{1}{3}}=12$,
it appears to be leveling off, and the choice of $N_{q}^{\nicefrac{1}{3}}=6$
yields a gap that is underestimating by $9$ meV. Likewise, as seen
in Figure \ref{fig:kgrid-conv-of-GaAs} (b), a linear fit of the last
three gaps as a function of $N_{q}^{-1}$ shows an intercept at 1.235
eV, suggesting the choice of a 6$\times$6$\times$6 $q$-grid underestimates
the bands gap by $12$ meV. From these considerations we conclude
that the choice of a 6$\times$6$\times$6 $q$-grid underestimates
the band gap by about 10 meV.

\newpage{}

\section{DFT Results}

\subsection{Band Gaps}

\begin{table}[h]
\begin{centering}
\setlength{\tabcolsep}{0.06 in} \renewcommand{\arraystretch}{1.2}%
\begin{tabular}{|c||ccc||cc|}
\hline 
 & PBE & PBE0 & HSE & Ref & ZPR\tabularnewline
\hline 
InSb & -0.45 & 0.83 & 0.28 & 0.19 & 0.17$^{\text a}$, 0.02$^{\text f}$\tabularnewline
InAs & -0.35 & 0.91 & 0.38 & 0.37 & 0.35$^{\text a}$, 0.02$^{\text f}$\tabularnewline
Ge & 0.08 & 1.25 & 0.69 & 0.71 & 0.66$^{\text b}$, 0.05$^{\text f}$\tabularnewline
GaSb & -0.12 & 1.22 & 0.69 & 0.76 & 0.73$^{\text a}$, 0.03$^{\text f}$\tabularnewline
Si & 0.55 & 1.77 & 1.12 & 1.18 & 1.12$^{\text b}$, 0.06$^{\text f}$\tabularnewline
InP & 0.69 & 2.09 & 1.48 & 1.40 & 1.35$^{\text a}$, 0.05$^{\text f}$\tabularnewline
GaAs & 0.40 & 1.86 & 1.27 & 1.47 & 1.42$^{\text a}$, 0.05$^{\text f}$\tabularnewline
AlSb & 0.99 & 2.14 & 1.52 & 1.65 & 1.61$^{\text a}$, 0.04$^{\text f}$\tabularnewline
AlAs & 1.33 & 2.61 & 1.96 & 2.20 & 2.16$^{\text a}$, 0.04$^{\text f}$\tabularnewline
GaP & 1.57 & 2.92 & 2.26 & 2.35 & 2.27$^{\text a}$, 0.08$^{\text f}$\tabularnewline
AlP & 1.56 & 2.91 & 2.24 & 2.51 & 2.49$^{\text a}$, 0.02$^{\text f}$\tabularnewline
C & 4.18 & 5.99 & 5.23 & 5.85 & 5.47$^{\text c}$, 0.38$^{\text g}$\tabularnewline
AlN & 4.22 & 6.27 & 5.57 & 6.52 & 6.14$^{\text a}$, 0.38$^{\text g}$\tabularnewline
MgO & 4.56 & 7.19 & 6.46 & 8.36 & 7.83$^{\text d}$, 0.53$^{\text h}$\tabularnewline
LiF & 9.08 & 12.14 & 11.40 & 15.35 & 14.20$^{\text e}$, 1.15$^{\text h}$\tabularnewline
\hline 
MAE & 1.50 & 0.70 & 0.58 &  & \tabularnewline
MSE & -1.50 & 0.081 & -0.56 &  & \tabularnewline
Max Error & -6.38 & -3.20 & -3.95 &  & \tabularnewline
\hline 
\end{tabular}
\par\end{centering}
\caption{Additional band gaps (in eV) at the DFT level for the various compounds
under study. At the bottom of the table are the MAE (mean absolute
error), MSE (mean signed error), and Max Error; all are in eV and
measured relative to the reported reference values, which are arrived
at by incorporating ZPR corrections into experimental band gap data.
Experimental results are arrived at via an analysis of optical absorption
spectroscopy data, where excitonic effects are taken into account
to arrive at the fundamental gap (See \citep{wingBandGapsCrystalline2021}
for details).\protect \\
References: a: \cite{vurgaftmanBandParametersIII2001a}, b: \cite{madelungSemiconductorsDataHandbook2004},
c: \cite{clarkIntrinsicEdgeAbsorption1964}, d: \cite{whitedExcitonThermoreflectanceMgO1973a},
e: \cite{piacentiniNewInterpretationFundamental1975}, f: \cite{cardonaIsotopeEffectsOptical2005a},
g: \cite{ponceTemperatureDependenceElectronic2015}, h: \cite{chenNonempiricalDielectricdependentHybrid2018b,neryQuasiparticlesPhononSatellites2018a}}

\end{table}
\newpage{}

\subsection{Valence Bandwidths}

\begin{table}[h]
\begin{centering}
\setlength{\tabcolsep}{0.06 in} \renewcommand{\arraystretch}{1.2}%
\begin{tabular}{|c|c||ccc||c|}
\hline 
\multicolumn{2}{|c||}{} & PBE & HSE & PBE0 & Expt.\tabularnewline
\hline 
InSb & $E_{\Gamma,8v}-E_{\Gamma,6v}$ & 10.97 & 12.05 & 12.17 & 11.7$^{\text a}$ (XPS), 10.8$^{\text a}$ (ARPES)\tabularnewline
InAs & $E_{\Gamma,8v}-E_{\Gamma,6v}$ & 11.95 & 13.14 & 13.26 & 12.3$^{\text a}$ (XPS)\tabularnewline
Ge & $E_{\Gamma,8v}-E_{\Gamma,6v}$ & 12.87 & 14.19 & 14.32 & 12.6$^{\text a}$ (XPS)\tabularnewline
GaSb & $E_{\Gamma,8v}-E_{\Gamma,6v}$ & 11.79 & 12.96 & 13.09 & 11.6$^{\text a}$ (XPS), 11.64$^{\text a}$ (ARPES)\tabularnewline
Si & $E_{\Gamma,8v}-E_{\Gamma,6v}$ & 11.98 & 13.30 & 13.40 & 12.5$^{\text a}$ (XPS)\tabularnewline
InP & $E_{\Gamma,8v}-E_{\Gamma,6v}$ & 11.53 & 12.74 & 12.83 & 11.0$^{\text a}$ (XPS), 11.4$^{\text a}$ (IPES)\tabularnewline
GaAs & $E_{\Gamma,8v}-E_{\Gamma,6v}$ & 12.85 & 14.12 & 14.22 & 13.8$^{\text a}$ (XPS), 13.1$^{\text a}$ (ARPES)\tabularnewline
AlSb & $E_{\Gamma,8v}-E_{\Gamma,6v}$ & 10.96 & 12.07 & 12.17 & \textbf{---}\tabularnewline
AlAs & $E_{\Gamma,8v}-E_{\Gamma,6v}$ & 11.97 & 13.19 & 13.28 & \textbf{---}\tabularnewline
GaP & $E_{\Gamma,8v}-E_{\Gamma,6v}$ & 12.52 & 13.81 & 13.90 & 12.5$^{\text a}$ (ARPES)\tabularnewline
AlP & $E_{\Gamma,8v}-E_{\Gamma,6v}$ & 11.52 & 12.76 & 12.83 & \textbf{---}\tabularnewline
C & $E_{\Gamma,8v}-E_{\Gamma,6v}$ & 21.50 & 23.7 & 23.74 & 21$^{\text a}$ (XPS)\tabularnewline
AlN & $E_{\Gamma,6v}-E_{\Gamma,3v}$ & 5.99 & 6.51 & 6.55 & \textbf{---}\tabularnewline
MgO & $E_{\Gamma,15v}-E_{L,1v}$ & 4.52 & 4.92 & 4.95 & 6.5$^{\text b}$ (XPS), 7$^{\text b}$ (XES) \tabularnewline
LiF & $E_{\Gamma,15v}-E_{X,4v}$ & 3.00 & 3.14 & 3.16 & 3.4$^{\text c}$ (XPS)\tabularnewline
\hline 
MAE &  & 0.59 & 1.25 & 1.17 & \tabularnewline
MSE &  & -0.31 & 0.92 & 0.83 & \tabularnewline
\hline 
\end{tabular}
\par\end{centering}
\caption{Valence bandwidths (in eV) at the DFT level for the compounds under
study. For zinc blende materials, the valence bandwidth is defined
as the maximal energy difference between the top four (excluding spin
degeneracy) valence bands. For the wurtzite and rock salt compounds,
the valence bandwidth is defined as the maximal energy difference
between the top three valence bands for LiF and MgO and the top six
valence bands for AlN. At the bottom of the table are the MAE and
MSE; all are in eV and calculated using the leftmost reported experimental
values. Experimental data are obtained via XPS, angle-resolved photo-emission
spectroscopy (ARPES), and X-ray emission spectroscopy (XES). Due to
a lack of quality data on the contributions of ZPR in these results,
we do not attempt to correct for such effects in our analysis.\protect \\
References: a: \cite{goldmannSubvolume1989b}, b: \cite{kowalczykElectronicStructureSrTiO31977},
c: \cite{roCharacterizationLiFUsing1992}}

\end{table}
\newpage{}

\subsection{$d$ Band Positions}

\begin{table}[h]
\begin{centering}
\setlength{\tabcolsep}{0.06 in} \renewcommand{\arraystretch}{1.2}%
\begin{tabular}{|c||ccc||c|}
\hline 
 & PBE & PBE0 & HSE & Expt.\tabularnewline
\hline 
InSb & 14.44 & 16.43 & 16.32 & 17.1$^{a}$, 16.98$^{b}$, 17.41$^{c}$\tabularnewline
InAs & 14.04 & 15.93 & 15.86 & 16.9$^{a}$, 17.40$^{c}$, 17.38$^{d}$\tabularnewline
Ge & 24.47 & 27.51 & 27.38 & 29.4$^{f}$\tabularnewline
GaSb & 15.04 & 14.83 & 17.62 & 18.8$^{a}$, 18.9$^{g}$\tabularnewline
InP & 13.73 & 15.56 & 15.48 & 17.1$^{a}$\tabularnewline
GaAs & 14.64 & 17.26 & 17.17 & 18.7$^{a}$, 18.7$^{b}$, 18.82$^{c}$\tabularnewline
AlSb & 27.86 & 30.14 & 30.06 & 31.15$^{e}$, 31.60$^{d}$\tabularnewline
AlAs & 34.18 & 37.12 & 37.05 & 39$^{h}$\tabularnewline
GaP & 14.28 & 16.81 & 16.73 & 18.6$^{a}$, 18.7$^{c}$\tabularnewline
\hline 
MAE & 3.80 & 1.70 & 1.47 & \tabularnewline
MSE & -3.80 & -1.70 & -1.47 & \tabularnewline
\hline 
\end{tabular}
\par\end{centering}
\caption{QP highest $d$ band positions at the DFT level for the various compounds
under study which contain $d$ electrons. At the bottom of the table
are the MAE and MSE; all are in eV and measured relative to the leftmost
reported experimental values. All experimental data are obtained via
X-ray photo-emission spectroscopy (XPS). Due to a lack of quality
data on the contributions of ZPR in these results, we do not attempt
to correct for such effects in our analysis.\protect \\
References: a: \cite{shevchikDensitiesValenceStates1974}, b: \cite{cardonaPhotoemissionGaAsInSb1972},
c: \cite{leyTotalValencebandDensities1974}, d: \cite{waldropEffectInterfaceComposition1992},
e: \cite{ehlersAngleresolvedphotoemissionStudyAlSb1989}, f: \cite{krautSemiconductorCorelevelValenceband1983},
g: \cite{gualtieriRayPhotoemissionCore1986}, h: \cite{okumuraPhotoemissionStudiesAlAs1987}}
\end{table}
\newpage{}

\section{SRSH and $G_{0}W_{0}$@SRSH Sensitivity}

\subsection{$\alpha$ and $\gamma$ variance for LiF and AlP}

\begin{figure}[h]
\begin{centering}
\includegraphics[scale=0.29]{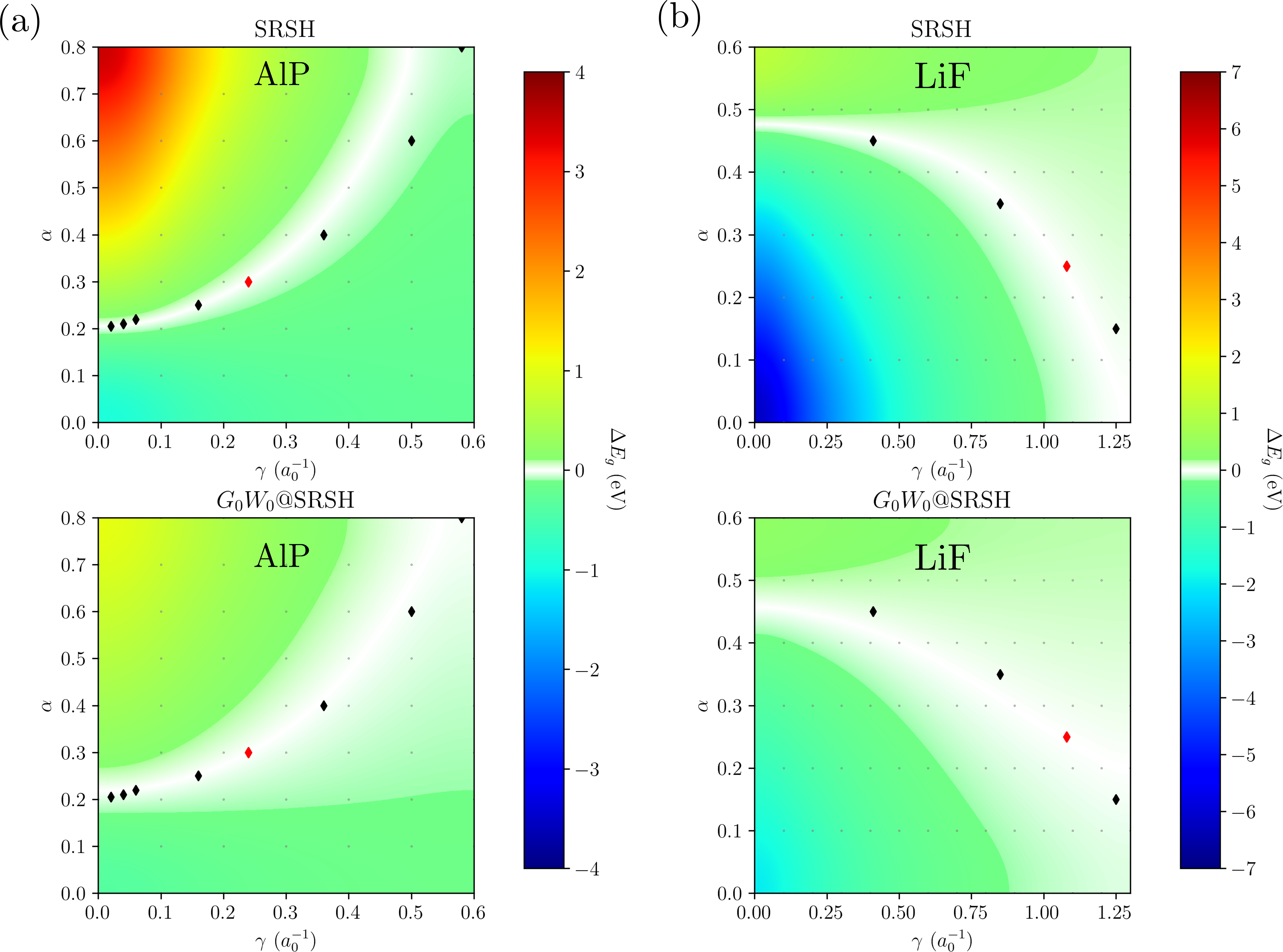}
\par\end{centering}
\caption{\label{fig:gap-sensitivity}Direct band gaps of AlP (a) and LiF (b)
relative to their reference value ($\Delta E_{g}=E_{g}-E_{\text{ref}}$)
at the SRSH and $G_{0}W_{0}$@SRSH levels, interpolated over a wide
range of $(\gamma,\alpha)$ values. The grid of performed calculations
is represented as gray dots, and the pairs satisfying the WOT-SRSH
constraint are depicted as diamonds, with the reference pair for $\Delta E_{g}$
in red. A range of $\pm100$ meV about the reference point is shaded
in white-green. $G_{0}W_{0}$ is seen to suppress the overall variation
at the SRSH level by about a factor of three.}
\end{figure}

As discussed in the main text, the application of $G_{0}W_{0}$ based
on a SRSH functional starting point suppresses how much the band gap
relative to a reference value (red diamond) $\Delta E_{g}$ changes.
Depicted in Figure \ref{fig:gap-sensitivity} are the interpolated
values of $\Delta E_{g}$ for AlP and LiF. Both compounds express
similar trends to those reported for AlN in the main text. The sensitivity
of $\Delta E_{g}$ is suppressed by about a factor of three, and the
optimal $\left(\alpha,\gamma\right)$ pairs (shown as black crosses)
tend to lie more tightly about the $\Delta E_{g}=0$ line for $G_{0}W_{0}$@SRSH.

One notable difference is that the optimal $\left(\alpha,\gamma\right)$
pairs for LiF have a negative slope, as opposed to AlN and AlP which
have series with positive slopes. A similar slope-change trend has
been observed in previous empirical tuning of SRSH functionals and
tends to occur for small $\varepsilon_{\infty}$, wide-band-gap materials
\citep{skoneNonempiricalRangeseparatedHybrid2016a}. 

\subsection{$\epsilon_{\infty}$ variance for AlN, AlSb, and LiF}

\begin{figure}[H]
\begin{centering}
\includegraphics[scale=0.4]{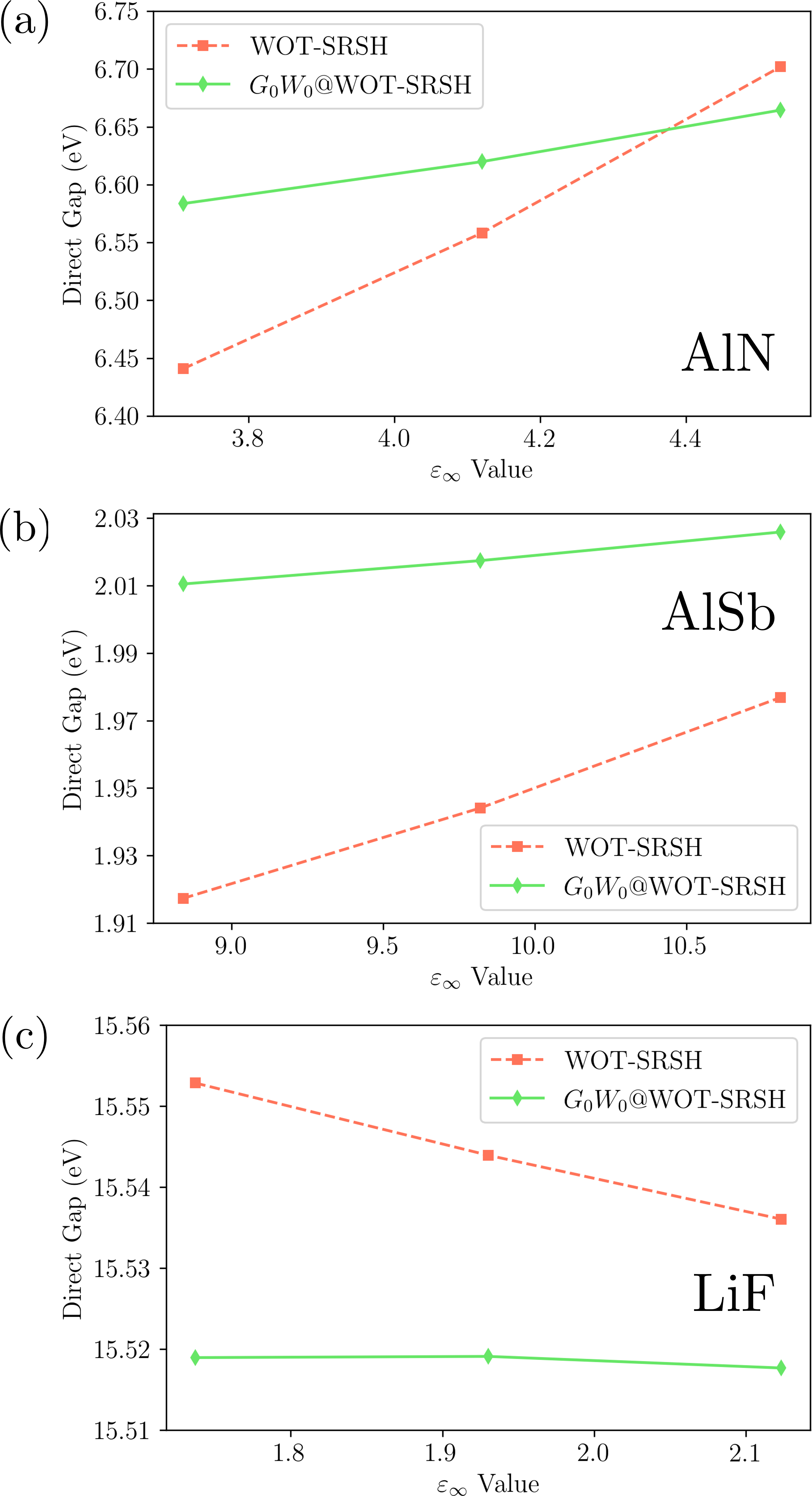}
\par\end{centering}
\caption{\label{fig:eps-sensitivity}Band gaps for various materials at the
SRSH and $G_{0}W_{0}$@SRSH levels as $\varepsilon_{\infty}$ is varied
by $\pm10\%$. $G_{0}W_{0}$ corrections suppress the variability
in the band gaps by a factor of three for AlN, AlSb, and substantially
more for than that for LiF.}
\end{figure}

\bibliographystyle{apsrev4-2}
\bibliography{refs}